# Nonequilibrium transport in epitaxial $CsPbBr_3$ single crystals

Seryio Saris[1], Roberto Rosati[2], Vladimir Bruevich[3], Thomas J. Sheehan[1], Maksim Román[1], Vitaly Podzorov[3], Ermin Malic[2], William A. Tisdale[1]

[1] Department of Chemical Engineering, Massachusetts Institute of Technology, Cambridge, MA, USA.

[2] Department of Physics, Philipps Universität Marburg, Marburg, Germany.

[3] Department of Physics and Astronomy, Rutgers University, Piscataway, NJ, USA.

## Abstract

Transport of optically excited carriers in semiconductors is typically described within a quasi-equilibrium picture, where energy is carried by a single thermalized quasiparticle population characterized by well-defined transport coefficients. Here, we demonstrate that in epitaxial $CsPbBr_3$ perovskite single crystals, this picture holds near room temperature – but breaks down dramatically at low temperature. Using transient microscopy, we show that optically measured carrier mobilities match device-scale Hall-effect and field-effect transistor measurements across a broad temperature range, resolving reported discrepancies and validating the equilibrium framework in the free-carrier regime. Below ~60 K, however, when excitonic effects become significant, equilibrium models begin to fail. We observe two coupled populations: a transient (<100 ps) hot-exciton gas with a diffusivity ~25–30 $cm^2/s$ – greatly exceeding the diffusivity expected for thermalized excitons – and a quasi-localized state that is fed by the cooling of the hot-exciton gas. These results reveal that in $CsPbBr_3$, transport and thermalization are not separable processes: carriers move while still redistributing among internal degrees of freedom, breaking the timescale separation that underpins equilibrium transport theory in conventional semiconductors. By resolving transport at the population level, we can directly access the competing kinetics of exciton formation, interconversion, and cooling, offering a new space for controlling energy flow in perovskite materials and their photonic applications.

## Introduction

Over the past decade, concerted theoretical and experimental efforts have converged on a broadly coherent microscopic picture of charge transport in bulk (3D) halide perovskites. Strong dielectric screening in the extended polar lattice yields photoexcited carriers that behave predominantly as independent electrons and holes, scattering frequently with phonons, defects, and other carriers.[1–6] Strong coupling of charge carriers to polar optical phonons via the Fröhlich interaction leads to the formation of large polarons,[7–10] in which charge is dressed in a lattice polarization cloud. While the microscopic details of carrier–phonon coupling in these soft, strongly anharmonic lattices remain debated,[11,12] the free-carrier Fröhlich polaron framework underpins much of our current understanding of carrier transport in halide perovskites.[13–18]

However, this picture rests on an implicit assumption that photoexcited carriers thermalize into a single, equilibrium population on a timescale much faster than that of spatiotemporal transport processes. In bulk halide perovskites, optical excitation generates not only free carriers but also bound electron–hole pairs (*i.e.* excitons) over a broad range of temperatures and carrier densities.[19–22] Importantly, the relative populations of these two species have been shown to deviate significantly from thermodynamic expectations based on the Saha equation.[23–25] These observations suggest that dynamic processes, including exciton and polaron formation, carrier cooling, and interconversion, are not necessarily "instantaneous" on transport-relevant timescales, and that carriers can move before full thermalization.[26,27] It therefore remains unclear how this richer, nonequilibrium, multispecies photophysical landscape shapes transport, and where the limits of the established free-carrier polaron framework lie.

The central obstacle to disentangling these effects is the lack of agreement in reported carrier mobilities across experimental probes. Measurements that access different timescales, such as steady-state electrical transport and transient optical techniques, often yield values that differ by orders of magnitude, even for nominally identical compositions.[17,28–32] Because each method investigates distinct spatiotemporal regimes and selectively samples different subpopulations of photocarriers, no consistent baseline for intrinsic mobility can be established. This inconsistency is further amplified by extrinsic factors such as disorder, grain boundaries, and structural heterogeneity, which can mask the underlying behavior.

Here, we address this challenge by mapping optically excited carrier transport in high-quality epitaxial $CsPbBr_3$ single crystals over a broad temperature range (5–300 K) using transient photoluminescence microscopy. At intermediate-to-high temperatures, the measured mobilities quantitatively match device-scale Hall-effect and field-effect transistor values, thereby establishing a unified benchmark across probes. Adding further confidence to our physical understanding, these experimental findings are also in quantitative agreement with the predictions of a microscopic transport model that treats free carriers and excitons as distinct channels within a Saha-equilibrium framework, validating the established picture of transport dominated by free-carrier polarons. Below ≈60 K, however, the measurements deviate strongly from the equilibrium transport model as excitonic effects become increasingly important. Rather than a single population, we resolve two coupled excitonic states: a fast, high-energy population and a slow, low-energy population. The high-energy population exhibits a "hot" Boltzmann distribution that cools within <100 ps, which tracks the anomalously fast spatial expansion both in timescale and temperature dependence. At the same time, population transfer to the localized low-energy state occurs within 15–20 ps. These observations reveal a nonequilibrium regime in which carrier motion is governed by the coupled evolution of excitonic populations rather than a fixed equilibrium species.

## Benchmarking temperature-dependent carrier transport

Epitaxial $CsPbBr_3$ single crystals were grown on mica via vapor-phase deposition, producing large millimeter-scale domains alongside smaller regions crystallographically aligned to the underlying substrate (Fig. 1a and Fig. S4). These crystals exhibit exceptional long-range order, high chemical purity and atomically flat surface morphology,[33] with minimal microscale defects and grain-boundary effects. High-resolution scanning transmission electron microscopy (STEM) confirms a well-ordered lattice across multiple regions of the sample, with a crystal structure consistent with the orthorhombic (*Pnma*) phase at room temperature (Extended Data Fig. 1). To further improve optical quality and minimize surface-environment interactions, the perovskite was encapsulated with a highly conformal insulating layer of parylene-N polymer (Extended Data Fig. 1).

We directly probed carrier transport in these epitaxial crystals using transient photoluminescence microscopy (TPLM), which is sensitive to both free carriers and excitons.[34–36] Here, a near-diffraction-limited pulse creates a localized carrier population that subsequently spreads away from the excitation spot, and a scanning avalanche photodiode records the time-dependent spatial evolution of the resulting photoluminescence (PL) profile (see Methods for details).

Representative temperature-dependent measurements are shown as spatiotemporal intensity maps in Fig. 1b. The PL profile expands over time, with pronounced broadening at lower temperatures. At the low excitation densities used here ($\leq 10^{16}$ $cm^{-3}$), carriers spread laterally faster than recombination can significantly reshape the emission profile, so the observed broadening is governed by carrier transport rather than effects from nonlinear decay processes (see Extended Data Fig. 2 and SI section 3.1).[37,38] To quantify the time-dependent spatial evolution, each PL profile is normalized and fit using a Gaussian-convolution model, taking the measured profile at $t = 0$ as the initial condition (see Methods for details). The variance $\sigma^2(t)$ is then extracted at each delay time, and the mean-squared displacement is defined as MSD = $\sigma^2(t) - \sigma^2(0)$, where diffusivity is obtained from the slope (MSD = $2Dt$).

Within the prompt-emission window, the MSD grows linearly, allowing us to extract a room-temperature diffusivity of $D = 0.92 \pm 0.04$ $cm^2$ $s^{-1}$. This value agrees with local measurements on single-crystal perovskites,[17,39,40] and is approximately an order of magnitude larger than typical diffusivities reported for polycrystalline thin films.[5,41,42] Earlier reports on $CsPbBr_3$ showed that nanoscale disorder, including local strain and point defects, can produce up to fourfold variations in diffusivity within single-crystal domains.[43] In contrast, our intradomain measurements show only minimal spatial variation (Fig. 1d), demonstrating that local inhomogeneity does not limit transport properties in our samples.

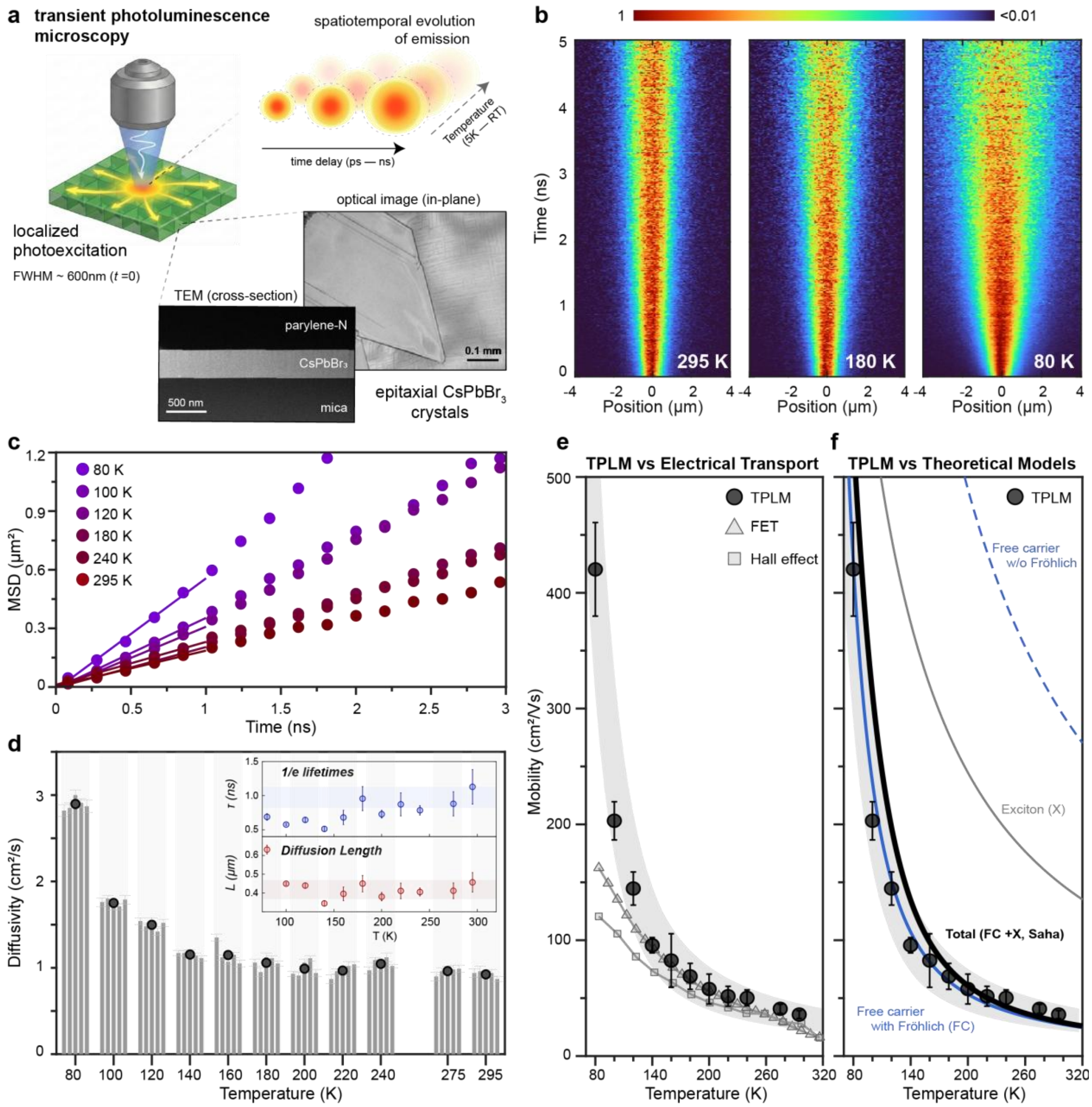

**Figure 1 | Benchmarking temperature-dependent carrier transport in epitaxial $CsPbBr_3$ perovskite.** **(a)** Transient photoluminescence microscopy (TPLM) scheme, an optical micrograph of the crystals showing millimeter-scale regions and smaller crystalline domains, and a cross-sectional TEM image of the sample **(b)** Temperature-dependent TPLM maps showing the evolution of the PL profile following localized excitation. **(c)** Corresponding mean-squared-displacement (MSD) curves extracted from (b), illustrating linear, diffusive growth. **(d)** Diffusivities extracted from the MSD slopes as a function of temperature; insets show the associated PL lifetimes and diffusion lengths ( $L = \sqrt{2D\tau}$ ). Each vertical bar represents measurements within a crystal domain **(e)** Comparison of TPLM-derived mobilities to Hall effect and FET measurements on identically grown samples [33] **(f)** Comparison of TPLM data to microscopic quasi-equilibrium transport models, showing calculated free-carrier (blue) and exciton (gray) mobility branches and their Saha-weighted total mobility (black) at carrier density 1 x $10^{15}$ $cm^{-3}$, similar to our excitation conditions. TPLM: 405nm, 0.19 μJ /$cm^2$ @ 5MHz.

The temperature dependence shown in Fig. 1d points to a carrier transport regime that is phonon-scattering limited. Diffusivity evolves nonlinearly upon cooling, remaining nearly constant from 300 K to ~140 K before rising sharply at lower temperatures. This overall trend is incompatible with a thermally activated hopping or defect-related scattering mechanism, which would lead to decreasing diffusivity at lower temperature. Interestingly, the prompt PL lifetime τ shows a modest overall decrease toward lower temperatures, resulting in a temperature-independent diffusion length of ~400 nm (Fig. 1d inset).

To determine whether the measured diffusivities and temperature trend reflect an intrinsic property of the material, we compare our TPLM-derived mobilities (via the Einstein relation, $D = \mu \frac{k_B T}{q}$) to Hall-effect and field-effect transistor (FET) measurements performed on identically grown $CsPbBr_3$ crystals from the same lab (Fig. 1e). Strikingly, the three techniques converge to similar numbers. From 300 K down to ~140 K, all yield closely matching values, with $\mu \approx 25$–$35\ cm^2/Vs$ at room temperature. This agreement is revealing, as each technique probes a fundamentally distinct transport regime: TPLM captures local, optically driven, transient carrier motion, whereas Hall-effect and FET mobilities are derived from macroscopic, steady-state electrical transport across device-scale dimensions. Furthermore, Hall-effect and FET measurements are sensitive only to free carriers, with no contribution expected from excitons. Their convergence with TPLM therefore demonstrates that, within this temperature range, all three probes access the same quasi-equilibrium carrier population. This provides model-independent evidence that free-carrier transport dominates above 140 K. However, at lower temperatures, device-based techniques begin to encounter extrinsic limitations, as electrical mobilities fall below the TPLM values. This divergence coincides with the emergence of cryogenic microcracking previously reported for vapor-grown $CsPbBr_3$ single crystals [33] and with structural features consistent with microcracks observed upon cooling (see Extended Data Fig. 3). Importantly, such cracking disrupts macroscopic conduction but does not measurably affect our local optical probe.

To understand the microscopic origin of this temperature-dependent behavior, we develop a phonon-limited transport model that treats free carriers and excitons as distinct species with fundamentally different coupling to the lattice. The model is constructed from first principles using electronic structure inputs from density functional theory, and with no adjustable parameters to scale the magnitude of the mobility (see SI section 1 for details). For free carriers, the dominant scattering channel is Fröhlich coupling to the $E_{LO} \approx 18$ meV longitudinal optical phonon (LO) mode of the Pb–Br sublattice,[44] and a weaker acoustic-phonon contribution extracted from the experimental PL linewidth (see SI section 4.2 and Extended Data Fig. 4). Excitons, in contrast, are assumed to have strongly suppressed Fröhlich coupling due to their vanishing net dipole, consistent with previous reports in bulk polar semiconductors.[45,46] The difference in scattering physics leads to two distinct mobility branches, as shown in Fig. 1f (solid blue and gray lines). The total diffusivity (Fig. 1f, solid black line) is then the weighted average of these two populations,[47] described by the Saha equation using the temperature-dependent exciton binding energy ($E_B \approx 19$ meV at T = 0 K, see Fig. S1).[44,48]

As shown in Fig. 1f, excitons are expected to play a negligible role in transport across the intermediate-to-high temperature range (~80-300 K), which is supported by the experimental findings. The close agreement between model predictions and experimental observations across this temperature regime is striking. In particular, the model captures a non-trivial temperature dependence of the scaling coefficient in $\mu \propto T^{-\alpha}$, where $\alpha$ increases upon cooling due to the suppression of Fröhlich scattering once thermal energy drops below $E_{LO}$ (see Fig. S2). The nonlinear temperature dependence of the diffusivity in Fig. 1d therefore emerges directly from this phonon-limited picture, with the sharp upturn at ~140 K marking the regime where LO phonons are no longer abundant. The concurrent decrease in lifetime (Fig.1d, inset) is also consistent with reduced polaronic screening, which increases the spatial electron–hole overlap and accelerates recombination.[49]

## Emergence of the excitonic regime below 60 K

The quantitative success of the free-carrier-polaron picture above ~80 K establishes the benchmark transport response of $CsPbBr_3$. Cooling further is expected to modify the phonon-scattering landscape, and by Saha arguments, shift the balance toward a higher excitonic fraction as thermal energy drops well below the binding energy. To investigate how these changes affect the emission spectrum, we track the temperature-dependent PL spectrum of our samples under low-fluence excitation (Fig. 2a). Near room temperature, both the PL resonance and linewidth reflect the anharmonic, strongly screened, and dynamically fluctuating lattice characteristic of lead-halide perovskites.[15,50] The peak position varies only weakly with temperature, and the homogeneous linewidth is dominated by LO-phonon–driven broadening (see Extended Data Fig. 4 and SI section 4.2).

Below 60 K, the system enters a distinct low-temperature regime. The PL peak red-shifts more steeply due to the positive deformation-potential contribution to the bandgap.[51,52] Additionally, the dominant LO-phonon scattering is suppressed and the PL linewidth collapses at its acoustic phonon limit (see Extended Data Fig. 4). As shown in Fig. 2b in detail, the PL spectrum splits into two emissive features separated by ~ 16 meV at 5 K, with a dominant low-energy (LE) resonance and a weaker high-energy (HE) shoulder. The spectra lack the long red tails characteristic of defect-related continuum emission. Importantly, samples measured without polymer encapsulation exhibit more variable low-temperature PL, with some grains showing multiple long-tailed emissive features and others displaying substantially broader spectra (see Fig. S7). The improved optical quality of the encapsulated crystals reduces this variability, allowing the underlying emissive structure to be resolved more reproducibly.

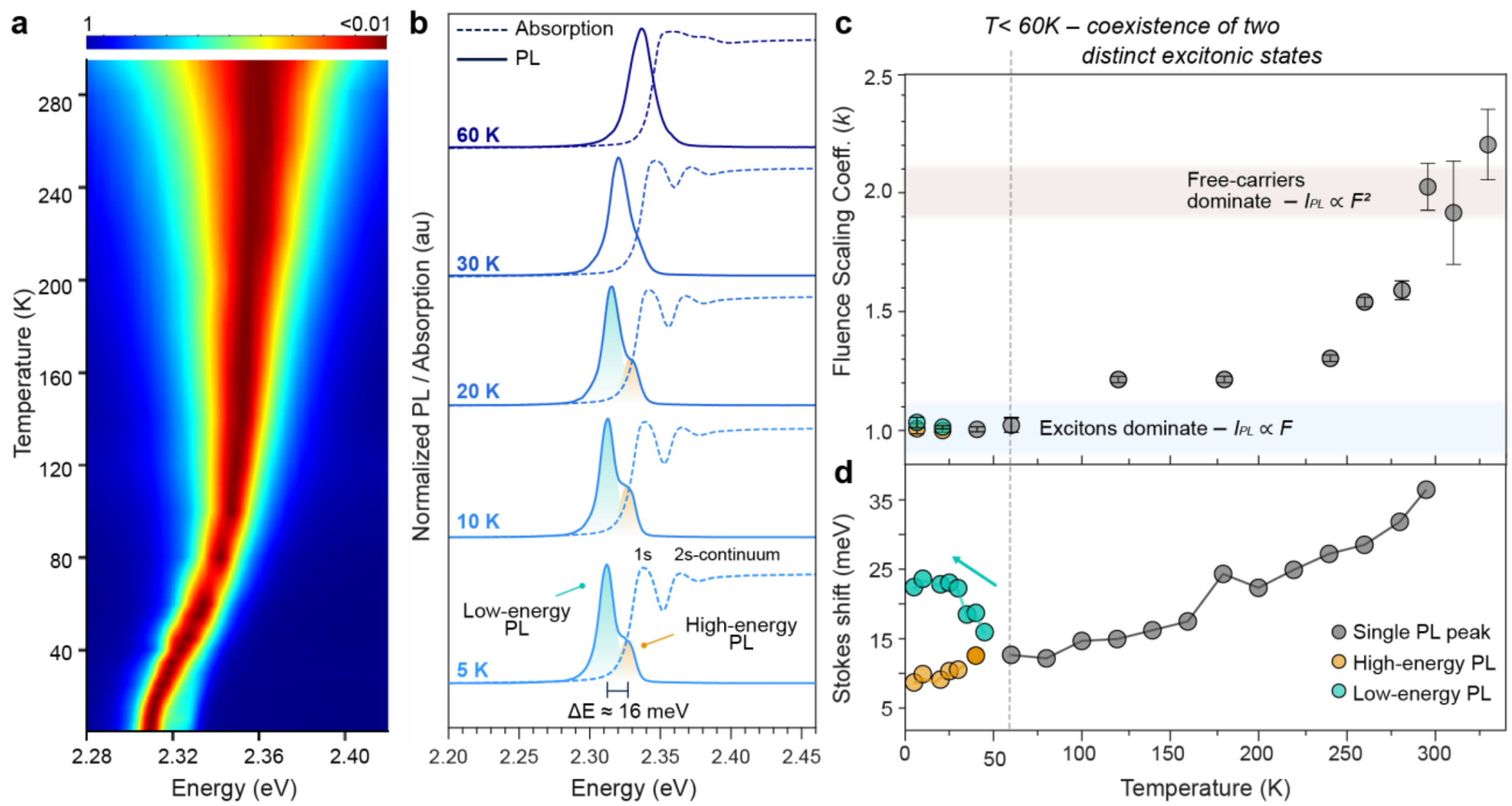

**Figure 2 | Emerging excitonic regime at low temperatures. (a)** Temperature-dependent PL map under pulsed non-resonant excitation at 5 MHz, normalized to highlight the evolution of the spectral shape **(b)** Normalized PL (solid) and absorption (dashed) spectra at selected temperatures. The single PL peak splits into two distinct features separated by ΔE ≈ 16 meV, assigned to high-energy (HE) and low-energy (LE) excitonic states. While the time-integrated spectra here show this splitting only below ~30 K, time-resolved spectral dynamics reveal that the two features persist up to 60 K (see Extended Data Fig. 6). **(c)** Scaling coefficient $k$ extracted from fluence dependence ($I \propto F^k$), showing a crossover from bimolecular free-carrier recombination at high temperature to a linear, exciton-dominated regime at lower temperatures. **(d)** Temperature-dependent Stokes shift. Below ~60 K, the LE emission deviates from the single-peak/HE trend and exhibits a pronounced upturn, consistent with a distinct low-energy excitonic state.

The two PL features could in principle arise from two excitonic states, or from the coexistence of excitons and free carriers. To establish a baseline description, we model the temperature-dependent PL spectra that include excitonic and continuum contributions based on the Saha equilibrium (see SI section 1.3 and Fig. S3). While this model captures the overall linewidth evolution well, it predicts a single excitonic peak to dominate below ~40 K, with a vanishing free-carrier contribution to the emission spectrum. The observed two-peak structure therefore cannot be explained by the equilibrium balance between excitons and free carriers at these temperatures.

Equilibrium thermodynamics alone, however, cannot exclude a kinetically stabilized free-carrier population persisting alongside excitons under pulsed excitation. To test this directly, we examine the fluence-dependent dynamics of the two features (Fig. 2c). The excitation fluence scaling behavior reflects the underlying recombination order,[53,54] where single-particle, excitonic emitters exhibit monomolecular scaling and uncorrelated electrons and holes lead to bimolecular behavior. Our measured scaling exponent $k$ in $I \propto F^k$ evolves clearly across the full temperature range. The PL scales quadratically near room temperature, consistent with free-carrier recombination[55,56] and our theoretical prediction (see Fig. S1), passes through a mixed regime upon cooling, and converges to a strictly monomolecular regime below 60 K. Notably, both HE and LE features exhibit excitonic scaling of $k \approx 1$ , with fluence-independent intensity ratios and peak positions (see Fig. S9).

Temperature-dependent absorption data reinforce the excitonic picture emerging at low temperatures. Sharp exciton peaks develop upon cooling (Fig. 2b and Extended Data Fig. 5), with the 1s exciton resonance closely tracking the HE PL feature. The Stokes shift decreases across the full temperature window (Fig. 2d), in agreement with previously reported behavior in $CsPbBr_3$ and other lead-halide perovskites.[50] Despite sharing excitonic character, the two peaks exhibit markedly different temperature dependences. As the PL spectrum splits into two components, the HE peak continues to follow the general Stokes shift trend and remains tied to the 1s resonance, whereas the LE feature diverges. This behavior argues against potential interpretations of a shared electronic origin of the emission peaks (*e.g.* phonon sideband or exciton fine-structure) and instead points to the HE and LE features being two distinct excitonic states.

## A breakdown of the quasi-equilibrium transport model

A natural question that follows is how carriers move in this emerging low-temperature regime where two excitonic populations coexist, and whether their spatiotemporal dynamics remain compatible with the quasi-equilibrium transport model that successfully captures the magnitude and temperature dependence of transport at higher temperatures.

As shown in Fig. 3a, spatially integrated transient PL decay accelerates markedly upon cooling, with the apparent lifetime decreasing by more than an order of magnitude from 60 K to 5 K. Such behavior is commonly observed in excitonic direct-gap semiconductors, where cooling increases the occupation of near-zero-momentum ($k \approx 0$) states, enhancing the fraction of excitons that can recombine optically.[57,58] Consistent with this picture, the lifetime broadly follows the expected $T^{3/2}$ scaling for thermalized free excitons in the bulk (Fig. 3a). However, below ~40 K, this scaling breaks down and an additional fast ~100 ps component emerges, coinciding with the increasingly pronounced HE–LE splitting in the PL (Fig. 2b). This behavior suggests the onset of two-population kinetics, in which distinct emissive populations contribute on different timescales, rather than a simple renormalization of a single radiative rate.

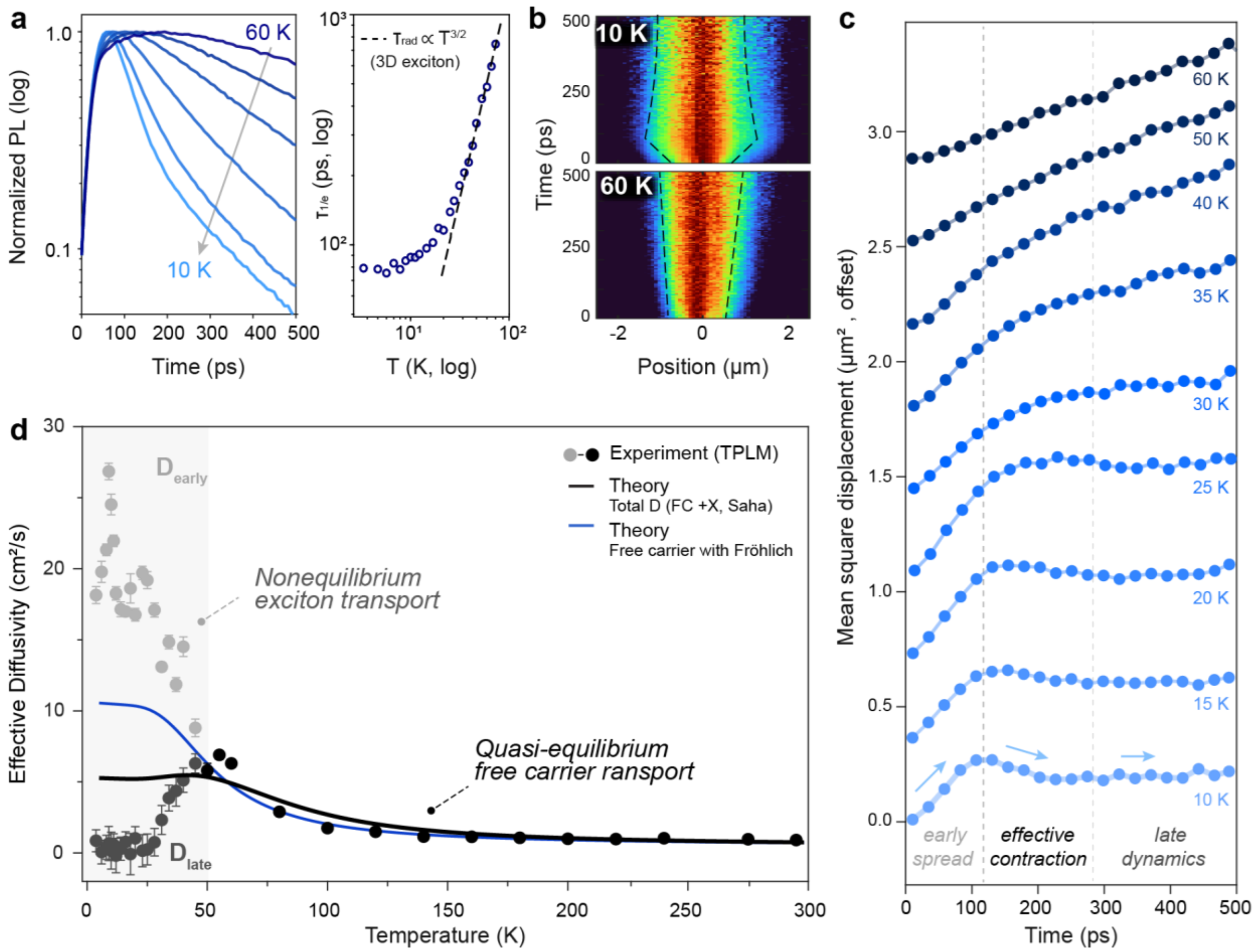

**Figure 3 | Nonequilibrium transport at low temperatures. (a)** Spatially integrated total PL dynamics and associated 1/e lifetimes. The dashed line shows the expected $\tau_{\mathrm{rad}} \propto T^{3/2}$scaling for thermalized 3D excitons for a comparison. **(b)** Spatiotemporal PL maps at 10 K and 60 K, showing distinct transport behavior, dashed lines here serve as guides to the eye. **(c)** MSD curves with strong non-linear dynamics, showing rapid early-time expansion, apparent contraction, and a slower, long-lived component at low temperatures. **(d)** Early-time (<100ps) and late-time (>250ps) effective diffusivities, $D_{\mathrm{early}}$ and $D_{\mathrm{late}}$, compared with the free carrier and Saha-weighted total quasi-equilibrium models. At this low-temperature regime both model predictions are limited by acoustic-phonon scattering. TPLM conditions: 405nm, 0.36 μJ /cm$^2$ @76 MHz.

The TPLM data show the same two-population signature in the transport dynamics (Fig. 3b–d). Below ~40 K, the MSD becomes strongly nonlinear, with a rapid expansion within the first ~100 ps followed by a pronounced slowing. At the lowest sample temperatures, the MSD even appears to contract at later times (Fig. 3c and Fig. S10), a behavior indicative of two diffusing populations with strongly contrasting diffusivities that are out-of-equilibrium with each other.[59–61] To track this nonlinear spatial evolution without imposing a single diffusion constant, we analyze the local MSD slope, which reveals a transient early-time high-diffusivity component followed by a collapse at later delays (see Fig. S11).

We therefore define two effective diffusivity metrics to quantify the time-dependent anomalous transport: an early-time peak diffusivity ($D_{\text{early}}$) and a late-time diffusivity ($D_{\text{late}}$), as plotted in Fig. 3d. Across this low-temperature regime, where optical-phonon scattering is frozen out, our equilibrium model predicts a saturation at the acoustic-scattering-limited value of ~5 $cm^2$/s (Fig. 3d, black line). Both $D_{\text{early}}$ and $D_{\text{late}}$ deviate sharply from this prediction. The measured $D_{\text{early}}$ reaches 25–30 $cm^2$/s — approximately 5–6× larger than the model prediction, and well above the upper bound set by the more conservative scenario of free carriers with smaller effective mass (Fig.3d, blue line). More strikingly, $D_{\text{early}}$ continues to rise as temperature decreases rather than saturating upon cooling. This inverse temperature trend is difficult to reconcile with a thermalized species whose transport is solely governed by the acoustic phonon bath (see SI section 8 for extended discussion).

In direct contrast to the early-time dynamics, $D_{\text{late}}$ collapses toward zero upon cooling the sample, with the fast channel giving way to a nearly immobile excitonic population. Its gradual increase upon warming is consistent with thermally activated escape out of a confining potential well into a more mobile state, which appears to occur at higher temperatures. Collectively, these two channels describe a system whose transport is fastest immediately after excitation and slows dramatically as the population evolves. Neither feature is compatible with a thermalized species in equilibrium with the lattice, and instead they call for a fundamentally nonequilibrium description of exciton motion.

## Nonequilibrium dynamics dominate early-time transport

To resolve the central mechanistic question of what drives nonequilibrium transport in $CsPbBr_3$ perovskites, we turn to the time-resolved spectral dynamics of the high-energy (HE) and low-energy (LE) emissive populations at low temperatures (Fig. 4 and Extended Data Fig. 6).

*Signature of hot-exciton dynamics:* We attribute the early-time fast transport to a transient population of hot excitons, which is observable through the time-dependent spectral shape of the HE emission peak (Fig. 4a,b). The emission spectrum is dominated by radiative recombination of excitonic states within the light cone; however, phonon-assisted recombination of transient hot excitons with finite momentum is also observable.[62,63] These hot excitons produce a continuous high-energy spectral tail whose slope reflects the underlying kinetic-energy distribution. Fitting this tail to a Boltzmann form therefore provides an effective measure of the excitonic temperature, $T_{eff}$ (see SI section 6 for details).

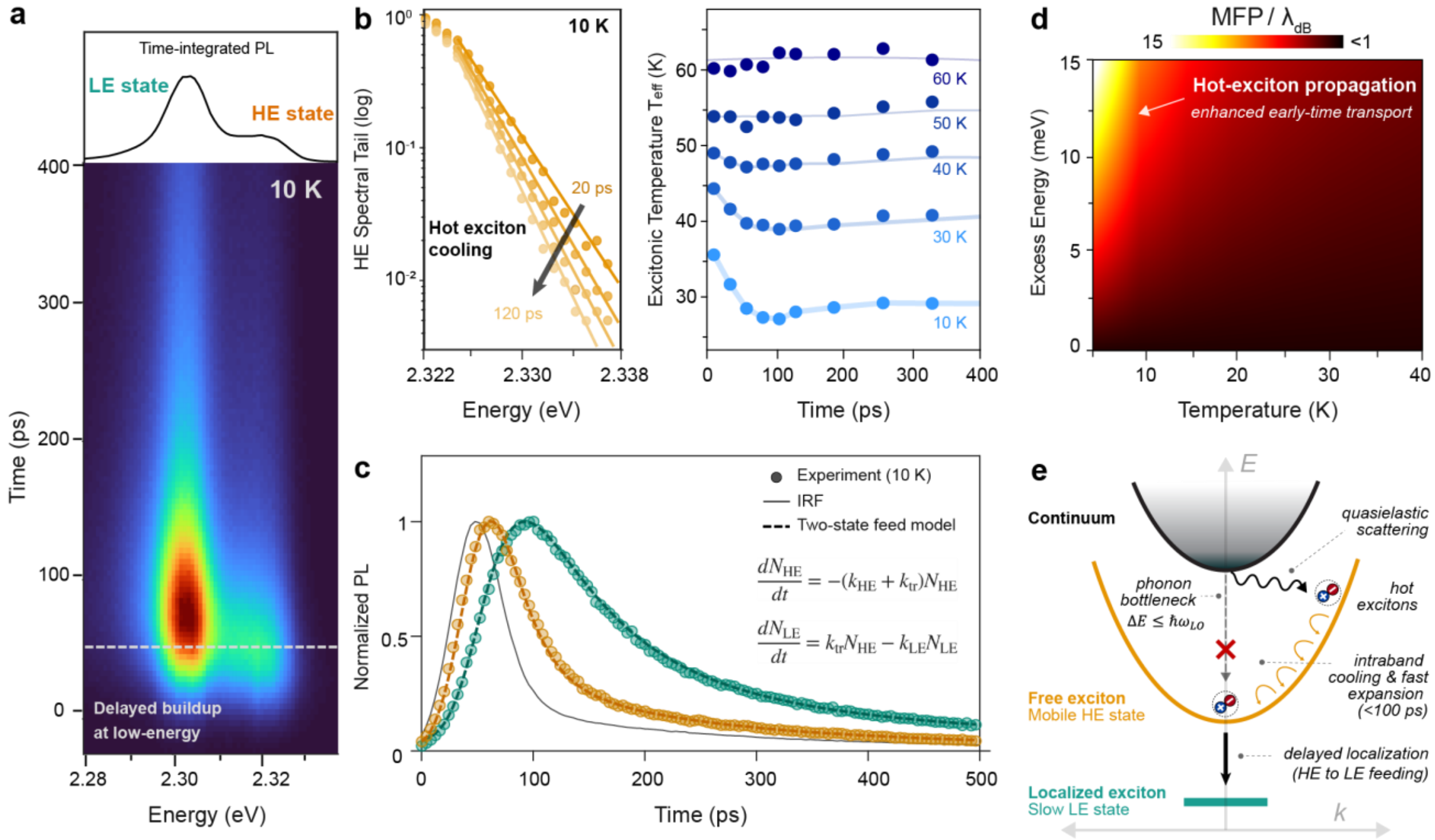

**Figure 4 | Nonequilibrium two-population dynamics and hot-exciton transport (a)** Spectrally resolved PL dynamics at 10 K showing that the high-energy (HE) channel decays rapidly while the low-energy (LE) channel exhibits a delayed buildup. Top panel shows the time-integrated PL spectrum. **(b)** Left: High-energy tail of the PL spectrum at selected delay times, showing a transient Boltzmann-like distribution. The slope steepens over time, indicating cooling of a hot exciton population on sub-100 ps timescales. Right: Extracted effective exciton temperature, $T_{eff}$, as a function of time for different lattice temperatures **(c)** Normalized PL dynamics of the HE and LE states (symbols), overlaid with a global two-state kinetic feed model (dashed lines) and the instrument response function (IRF, gray). The dynamics are well described by a transfer from the HE to LE population with a timescale of ~15 ps at 10 K **(d)** Excitons with finite excess energy lie in a semiclassical regime where the mean free path greatly exceeds the de Broglie wavelength and the quantum localization limit (MFP / λdeBr ≫1). **(e)** Schematic of the proposed hot-exciton transport pathway. Measurements at 405nm, 0.36 μJ /cm$^2$ @76 MHz.

At 10 K, we directly observe the spectral fingerprint and cooling dynamics of hot excitons in the HE tail (Fig. 4b). The initially elevated $T_{eff}$ relaxes toward a quasi-equilibrium value within ≤100 ps, ultimately limited by the finite linewidth of the HE peak (see SI section 6.2). At higher lattice temperatures, the cooling signature weakens and becomes indistinguishable from equilibrium with the lattice temperature above 40 K.

Crucially, this spectral relaxation mirrors the transport dynamics in Fig. 3c. Both observables relax on the same 100 ps timescale and both vanish over the same temperature range. Thus, the HE-tail thermometry and time-resolved transport measurements provide two independent probes of the same nonequilibrium population, linking the hot exciton states directly to the fast early-time spatial expansion. Transport and energy relaxation are consequently inseparable in this regime: rather than being characterized by a single equilibrium diffusivity, the apparent transport coefficient evolves in time together with the kinetic-energy distribution of the excitons themselves.

*Microscopic origin of hot excitons:* Any microscopic mechanism underlying the persistence of a hot exciton transport channel at low temperature must explain why the excitonic population fails to fully thermalize within ≤100 ps. We propose that the persistent hot-exciton population arises from a kinetic bottleneck in scattering from the free-carrier continuum into the free exciton manifold (Fig. 4e). In this picture, non-resonant photoexcitation generates free carriers that rapidly cool to the continuum band edge**.** At 10 K, however, further relaxation of electron-hole pairs into excitonic states becomes inefficient. Inelastic scattering channels are strongly suppressed[64], and direct single-step relaxation into the lowest-energy excitonic state via LO phonons can be kinematically constrained.[65,66] In $CsPbBr_3$, this constraint is particularly relevant given that the dominant LO phonon energy ($E_{LO} \approx 18$ meV*)* becomes comparable to the low-temperature binding energy ($E_B \approx$ 19 meV), leaving only a narrow subset of states able to satisfy energy and momentum conservation. Under these conditions, exciton formation proceeds predominantly through quasielastic acoustic-phonon scattering into high-$k$ excitonic states, generating excitons with excess kinetic energies on the order of $E_B$.

Simulation results show that a phonon-bottleneck-mediated formation of high-$k$ excitons can explain the anomalous transport trends observed experimentally. Unlike a thermalized excitonic population near the band minimum, hot excitons retain substantial excess kinetic energy and propagate with larger group velocities. From the standpoint of transport theory, the key finding is that these hot excitons have a mean free path that greatly exceeds the de Broglie wavelength (MFP/$\lambda_{dB} \gg 1$) in $CsPbBr_3$ at low temperature (Fig. 4d and Extended Data Fig. 7). In this regime, excitons traverse substantial distances before scattering and avoid localization during the early stages of transport. Our kinetic-theory estimates for transient diffusivities show that finite excess energy reproduces the experimentally observed increase in transient diffusivity upon cooling (Extended Data Fig. 7), a trend absent under equilibrium conditions and characteristic of hot-exciton transport. While excess energies comparable to $E_b$ provide an upper-bound scenario, the measured diffusivities are more consistent with smaller effective excess energies of a few meV, likely reflecting partial cooling during exciton formation and the finite instrument response (see SI section 8).

The same phonon-bottleneck-mediated hot-exciton formation picture also accounts for the disappearance of this behavior at higher temperatures. As LO-phonon-assisted relaxation becomes more efficient, hot excitons thermalize faster and the bottleneck is relieved. Above ~40 K, this cooling likely proceeds faster than the temporal resolution of our TPLM setup. However, other ultrafast measurements on $CsPbBr_3$ single crystals have reported orders of magnitude faster (~0.2 ps) cooling dynamics near 80 K , where efficient LO-phonon-mediated relaxation is expected.[67]

*Population transfer and localization:* A complete description of the nonequilibrium transport must also account for where the hot excitons end up once they have lost their excess energy after the first 100 ps. The transient spectral dynamics clearly reveal that the two emissive features are kinetically coupled (Fig. 4a,c). The HE peak rises promptly within the instrument response, whereas the LE peak builds up with an observable delay. A two-state kinetic model in which a promptly formed HE population feeds the LE state quantitatively reproduces the global spectral dynamics, yielding a feeding time of ~15 ps at 10 K (dashed lines, Fig. 4c; see SI section 7 for details). In the limit of negligible nonradiative recombination, we estimate that 75% of the HE population transfers into the LE state (see SI section 7.2). The HE state therefore acts primarily as a “hot” transient reservoir whose apparent decay is governed by population transfer rather than direct radiative recombination. The apparent feeding time lengthens with increasing temperature as HE→LE back-transfer grows comparable to the forward rate, and the two states approach thermal equilibrium (Table S1).

The population transfer into the LE manifold also explains the crossover from fast early-time transport to the slower dynamics observed after the first ~100 ps. As the hot HE excitons spread outward and lose their excess energy, they progressively transfer into the LE state and become pinned near where they were created. Hence, the apparent MSD contraction at the lowest temperatures (Fig. 3c and Fig. S10) is the fingerprint of this handoff, marking the point at which the slow LE state begins to dominate the emission. The LE population itself exhibits signs of a localized state, with thermally activated hopping dynamics and an activation energy of ~6–10 meV (see Fig. S12 and SI section 7.1), consistent with a shallow localization potential rather than a deeply self-trapped excitonic state.[68] The increased low-temperature Stokes shift of the LE resonance (Fig. 2d) further supports this assignment, as such shifts commonly arise from localized emissive states.[69]

While the exact origin of the LE localized state remains open, one possibility is that quasi-static structural disorder at low temperature creates shallow confinement sites for excitons. At cryogenic temperatures, the lattice dynamics in perovskites are slow and acquire a quasi-static character, giving rise to “frozen” octahedral tilts and distortions.[70] Recent experimental and theoretical studies[71,72] on bulk $CsPbBr_3$ report low-temperature exciton localization associated with shallow potential variations and the formation of polarization nanodomains, which support phonon-assisted hopping with characteristic energies set by soft optical modes on the order of ~7 meV—closely matching the activation energy extracted here.

## Outlook

In conventional bulk semiconductors, transport is typically described under the assumption of strong timescale separation, where nonequilibrium carrier populations relax within sub-ps timescales and transport proceeds within a fixed, effectively equilibrated energy landscape.[73,74] Our results show that this picture breaks down in epitaxial $CsPbBr_3$ crystals. Here, carriers do not first equilibrate and then move; instead, they move while redistributing between hot, mobile excitonic states and lower-energy localized states. As a result, the measured diffusivity is not a static material parameter, but a time-dependent response governed by the instantaneous population landscape. Population-resolved transport therefore exposes control variables that are invisible in equilibrium measurements, including the initial carrier energy distribution, exciton-formation pathway, branching between mobile and localized states, and cooling bottlenecks that connect them. This makes nonequilibrium transport not simply a deviation from the equilibrium limit but a route to controlling energy flow through the internal dynamics of the photoexcited population itself.

Within our specific system, an important question is what sets the lifetime of these hot, nonequilibrium populations, and whether it can be engineered. The <100 ps relaxation timescale appears to be governed by phonon-assisted cooling into the lower excitonic manifold, a process that could in principle be tuned through composition, dimensionality, or confinement. Extending this nonequilibrium window may open access to phenomena that equilibrium transport cannot reach: memory effects in which prior excitation history shapes subsequent transport,[75] or collective behavior, such as superradiance or superfluidity,[76,77] at carrier densities where correlations develop before relaxation. More broadly, systems in which transport and relaxation occur on comparable timescales offer a rich space for future study, with lattice dynamics, dimensionality, and disorder serving as tuning knobs for energy flow and material functionality.

## Extended data

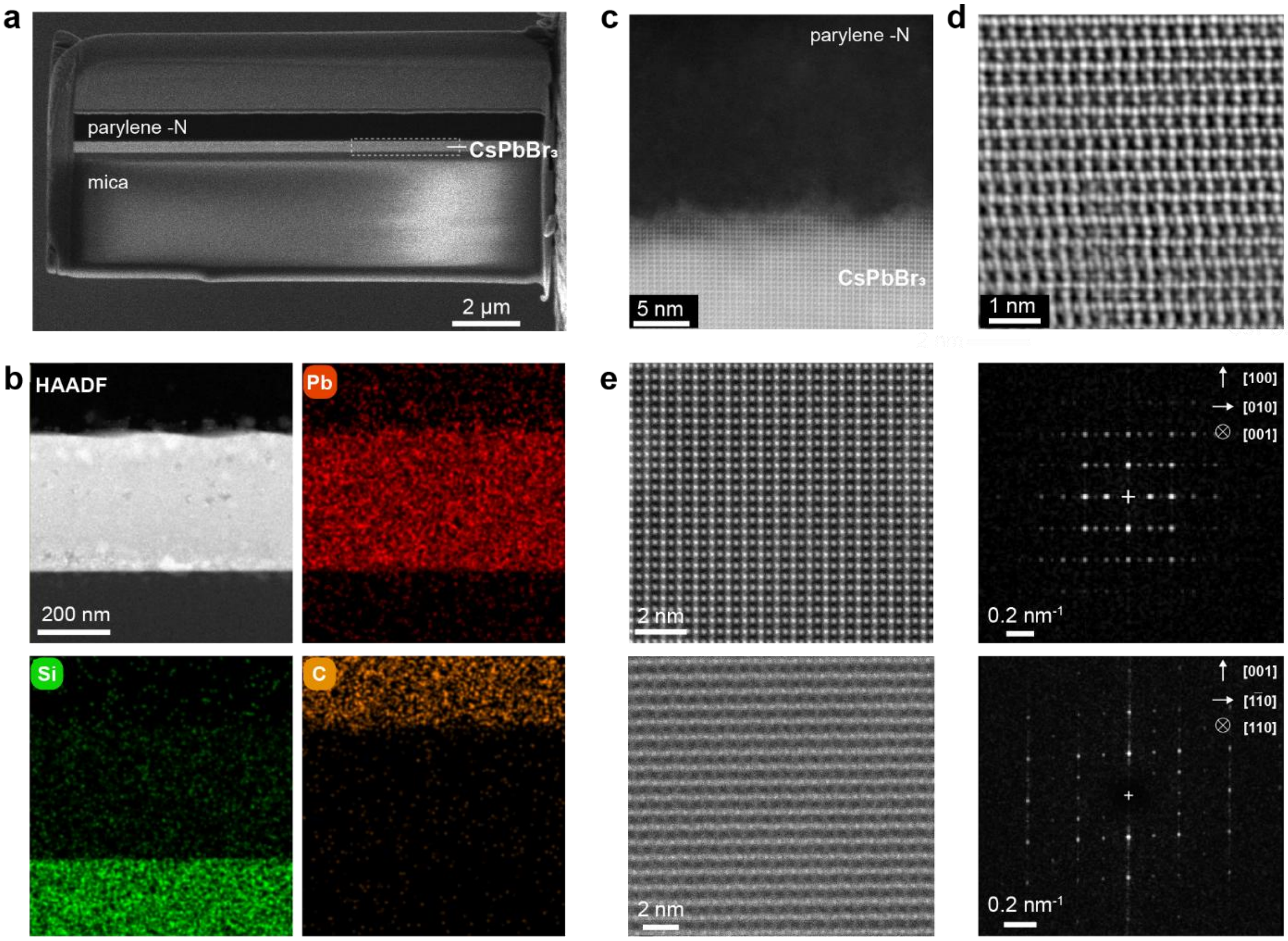

**Extended Data Fig. 1 | Structural and compositional characterization of $CsPbBr_3$ epitaxial crystals. (a)** Cross-sectional scanning electron microscopy image of a $CsPbBr_3$ film (~400 nm) grown on mica and encapsulated with a parylene-N polymer layer (~ 1 µm). The dashed region shows the interfacial region selected for high-resolution analysis. **(b)** High-angle annular dark field (HAADF) STEM image of the interface with corresponding elemental maps showing the spatial distribution of Pb (red), Si (green), and C (orange), confirming the layered structure of the samples. (**c)** High-resolution HAADF-STEM image of the $CsPbBr_3$/parylene-N interface, showing a highly conformal coating. **(d)** Atomic-resolution differential phase contrast (iDPC-STEM) images near the [110] zone axis. The phase contrast reveals additional features between Pb columns that are consistent with contributions from lighter elements**. (e)** Atomic-resolution HAADF-STEM images recorded along two crystallographic zone axes (left), with corresponding FFTs (right). The Pb sublattice is resolved with strong Z-contrast, and the diffraction patterns reveal anisotropic spot spacings along orthogonal directions. The observed splitting of reflections that would be equivalent in a cubic lattice, together with the reduced symmetry of the FFT patterns, supports assignment to an orthorhombic perovskite structure.

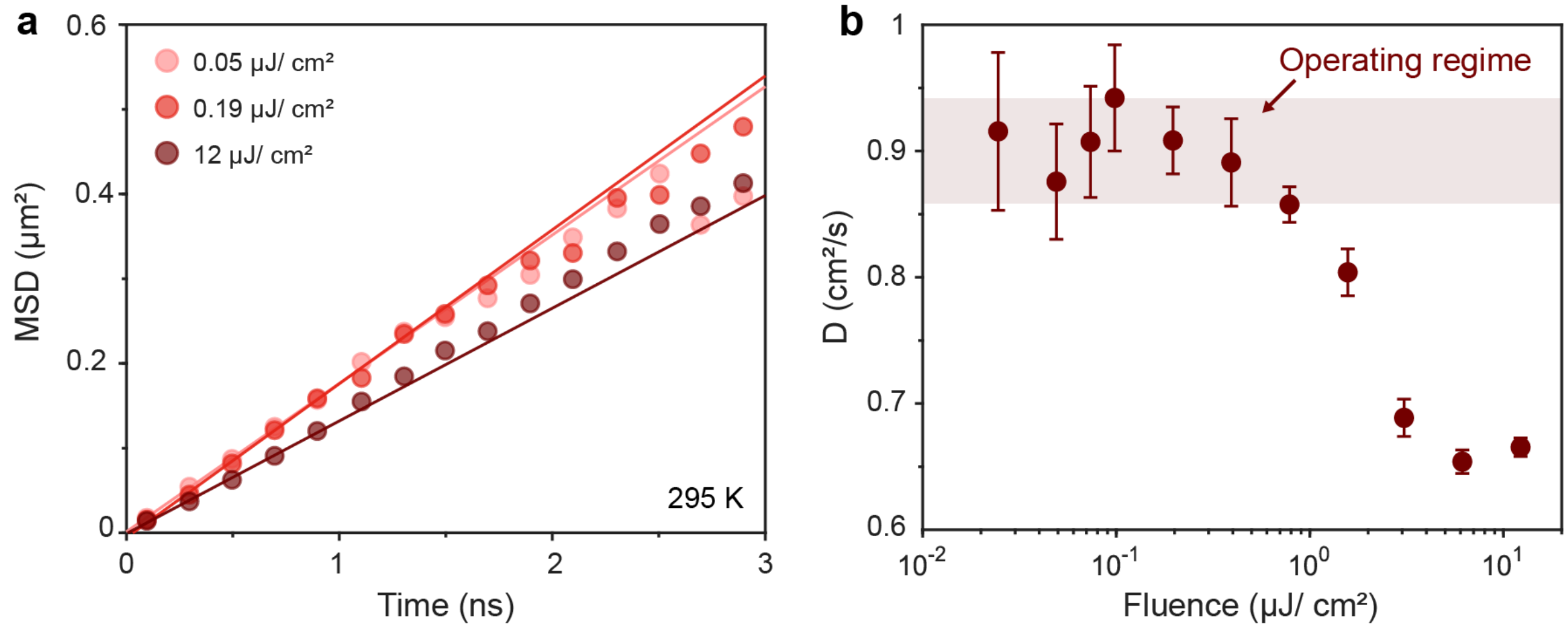

**Extended Data Fig. 2 | Fluence-dependent spatial broadening and diffusivity. (a)** Representative mean square displacement (MSD) traces at room temperature acquired under various excitation fluences. Linear fits to the MSD evolution exhibit comparable slopes at low to intermediate fluences. **(b)** Diffusivities extracted from first-order diffusion analysis as a function of excitation fluence. The excitation conditions employed in our main measurements lie within a regime where the extracted diffusivities remain nearly fluence independent. The fluence range probed corresponds to estimated average carrier densities spanning $3 \times 10^{14} – 3 \times 10^{17}$ $cm^{-3}$.

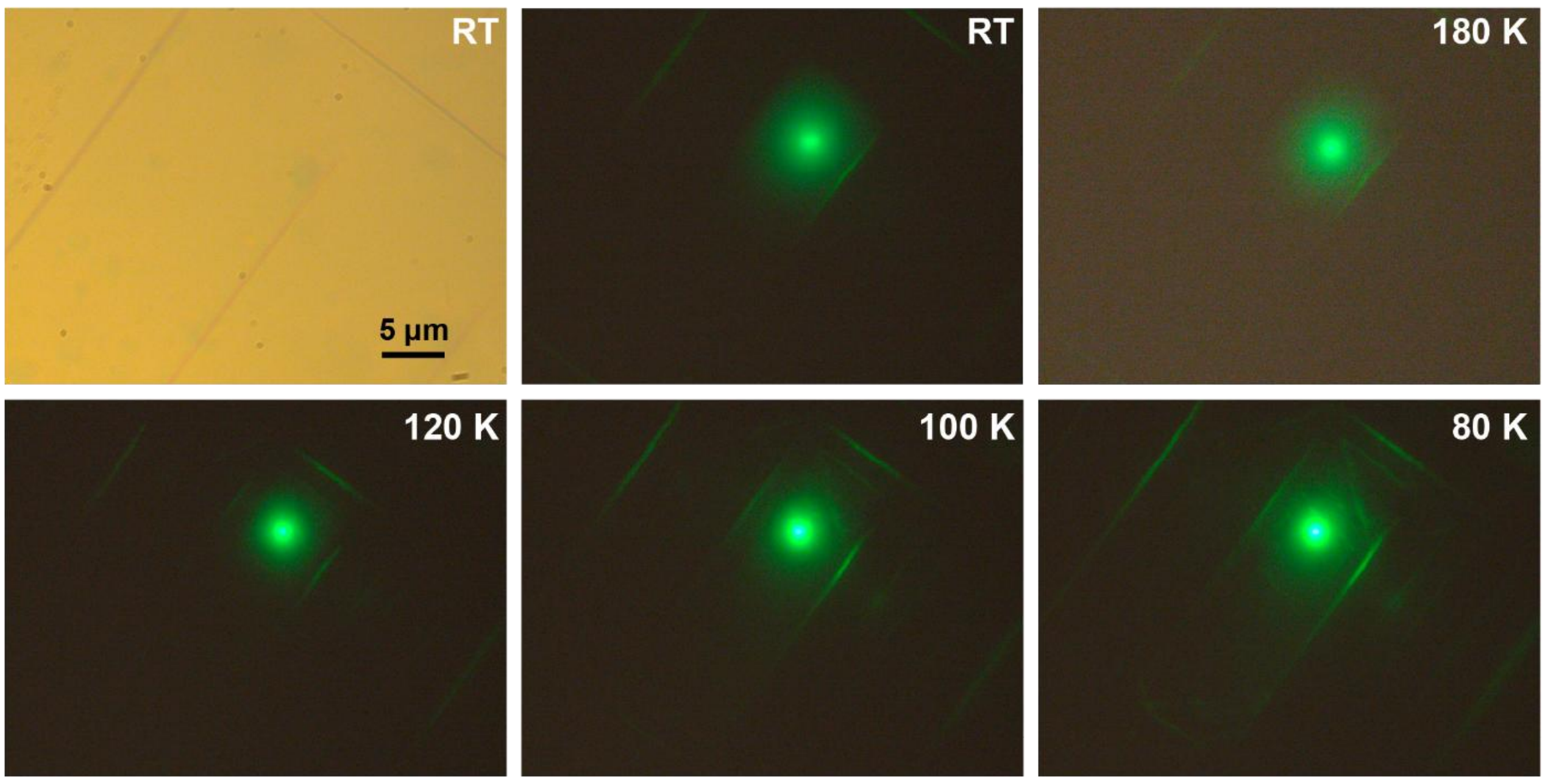

**Extended Data Fig. 3 | Temperature-dependent optical images of epitaxial $CsPbBr_3$ single crystals.** Images acquired between room temperature and 80 K reveal increasing structural contrast upon cooling, including diagonal line features consistent with thermally induced strain or microcrack formation. The excitation region is visible as the central green emission spot. Such microcracks have been previously shown to introduce discontinuities in macroscopic conduction pathways and affect electrical mobility measurements at low temperatures. For a detailed discussion on cryogenic microcracking in perovskites and FET/Hall-effect experiments, see Ref. [33].

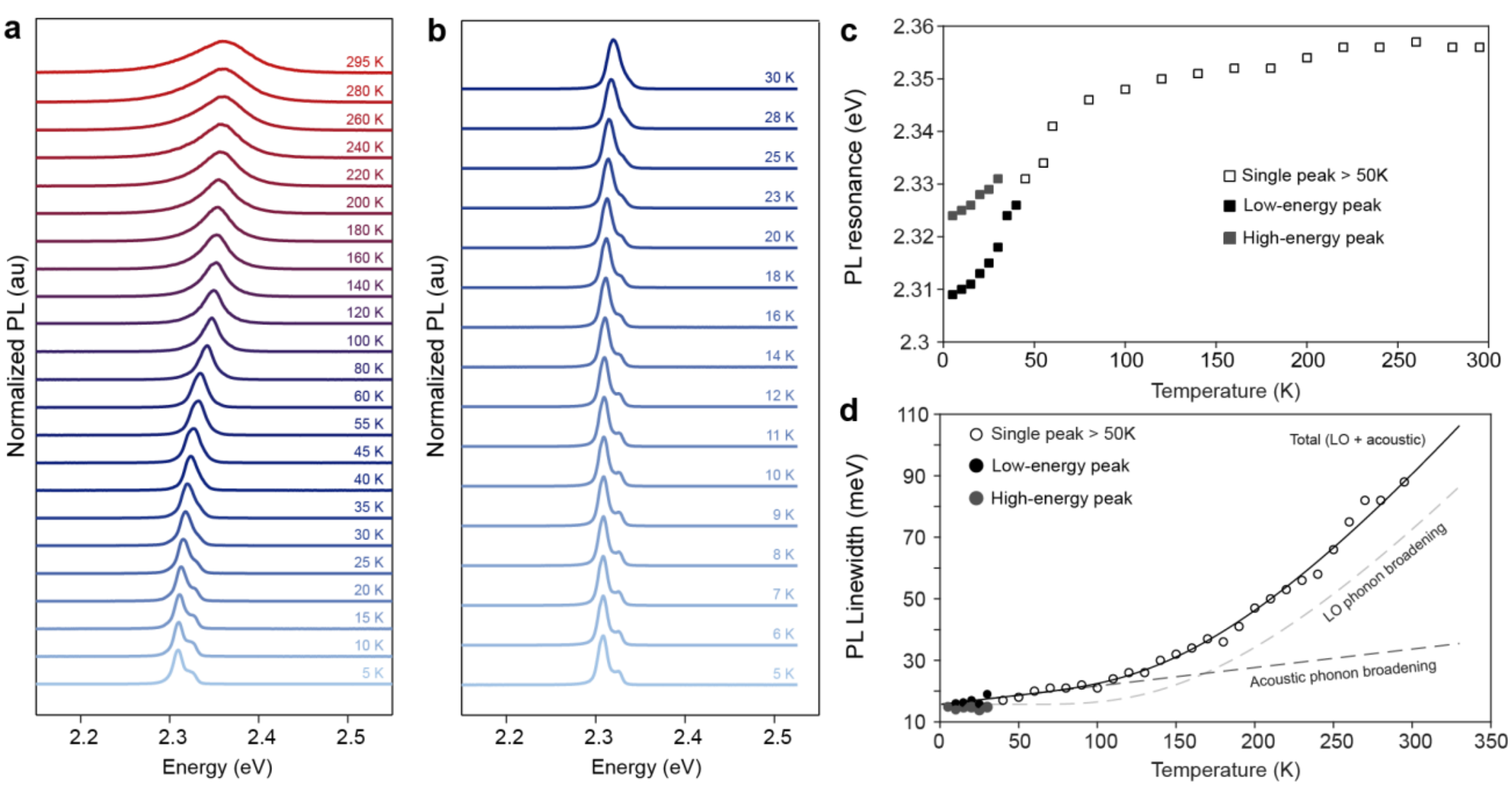

**Extended Data Fig. 4 | Temperature-dependent PL spectra and broadening of epitaxial $CsPbBr_3$ single crystals. (a)** Spectral evolution over the full temperature range (5–300K) **(b)** Expanded view of the low-temperature regime (5–30K), showing the emergence of two emissive populations. **(c)** Temperature dependence of PL resonance energy showing the characteristic red-shifting behavior in perovskites upon cooling **(d)** Temperature dependence of the PL linewidth, extracted as the full width at half maximum (FWHM). The solid line shows a phenomenological fit incorporating temperature-independent inhomogeneous broadening, which arises from intrinsic static disorder, together with acoustic and optical phonon scattering contributions to the total linewidth (dashed lines). For further details on the analysis of temperature-dependent PL broadening, see SI section 4.2. Experimental conditions: 405nm, 0.19 μJ /cm$^2$ @ 5MHz.

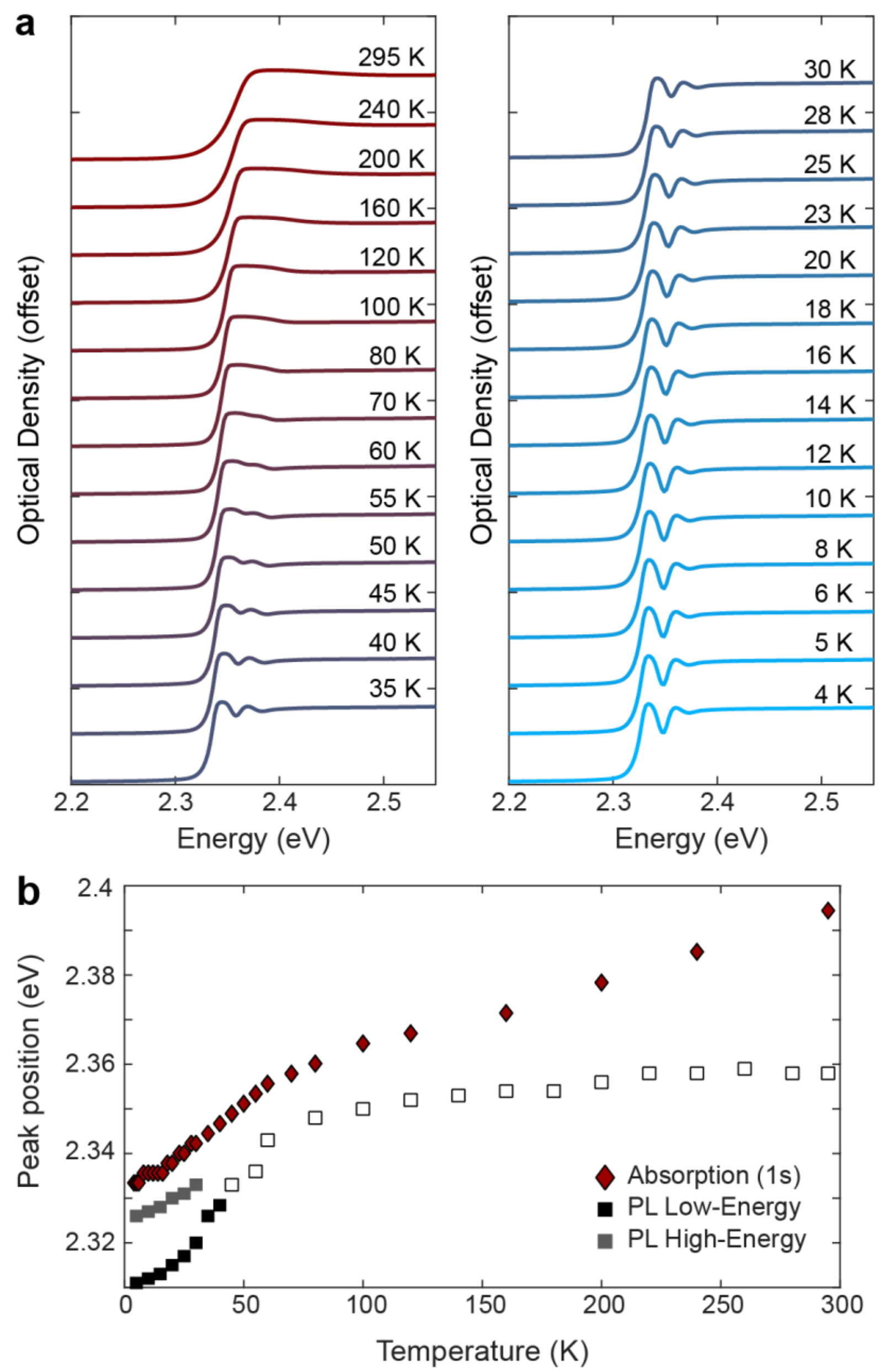

**Extended Data Fig. 5 | Temperature-dependent absorption spectra of epitaxial $CsPbBr_3$ single crystals. (a)** Offset absorption spectra over the full temperature range (4 K – RT), showing the sharpening at low temperatures and progressive broadening and blueshift upon warming. **(b)** Extracted peak positions as a function of temperature for the absorption (1s exciton) and photoluminescence peak energies.

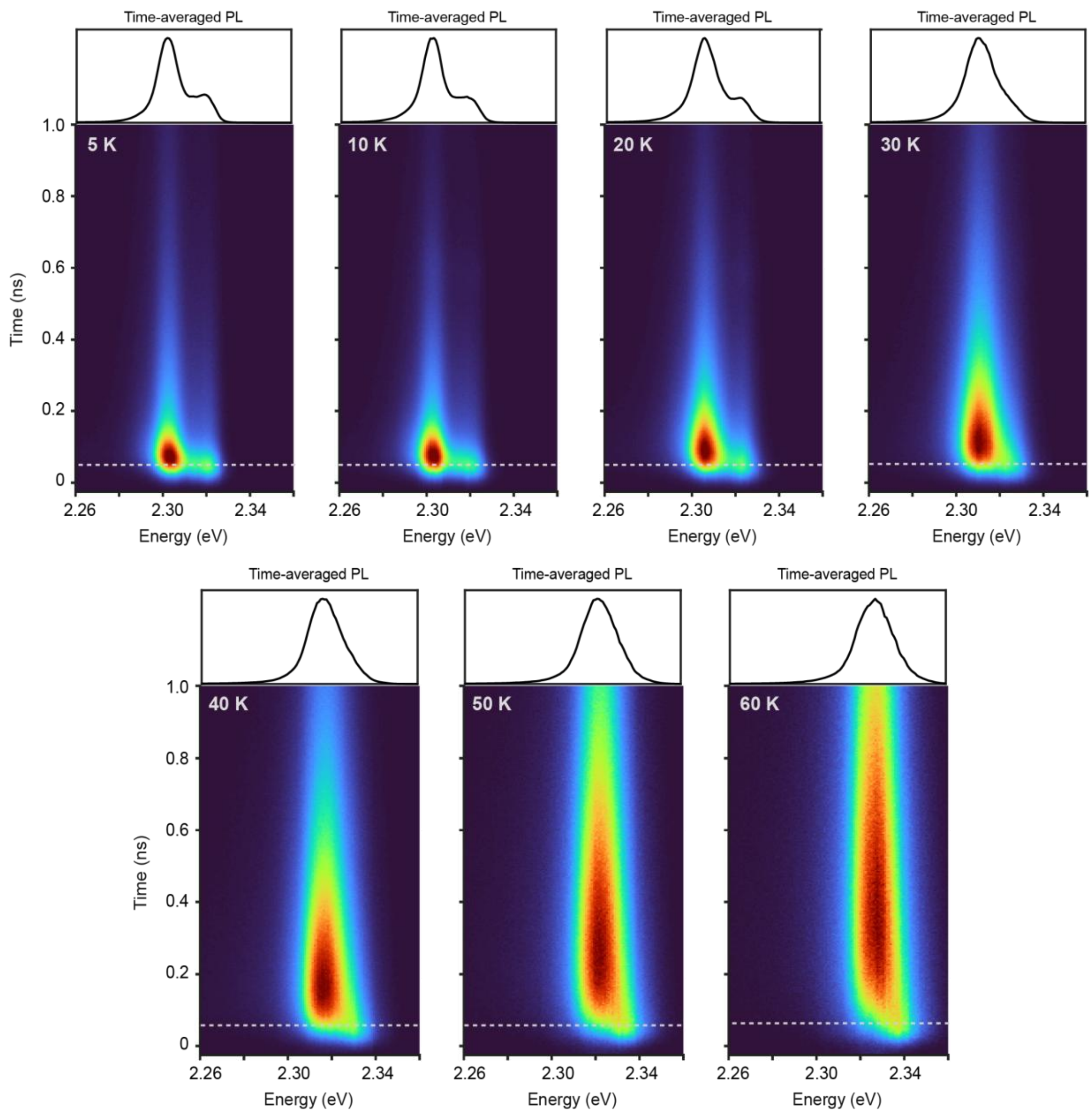

**Extended Data Fig. 6 | Temperature-dependent spectral dynamics in the low-temperature regime.** Time-resolved photoluminescence (PL) spectral maps measured from 5 K to 60 K, shown as energy–time colormaps. The top panels display the corresponding time-averaged PL spectra integrated over a 1 ns decay window. At all temperatures, the low-energy (LE) peak emerges with a clear delay relative to the high-energy (HE) emission (white dashed line indicates the peak of the HE emission), consistent with population transfer from HE into LE states. A distinct HE contribution remains visible in the early-time dynamics even at higher temperatures, where the time-averaged PL appears single-peaked. As the temperature increases from 5 K to 60 K, the LE component exhibits progressively longer lifetimes, while the HE–LE splitting becomes comparable to the thermal broadening.

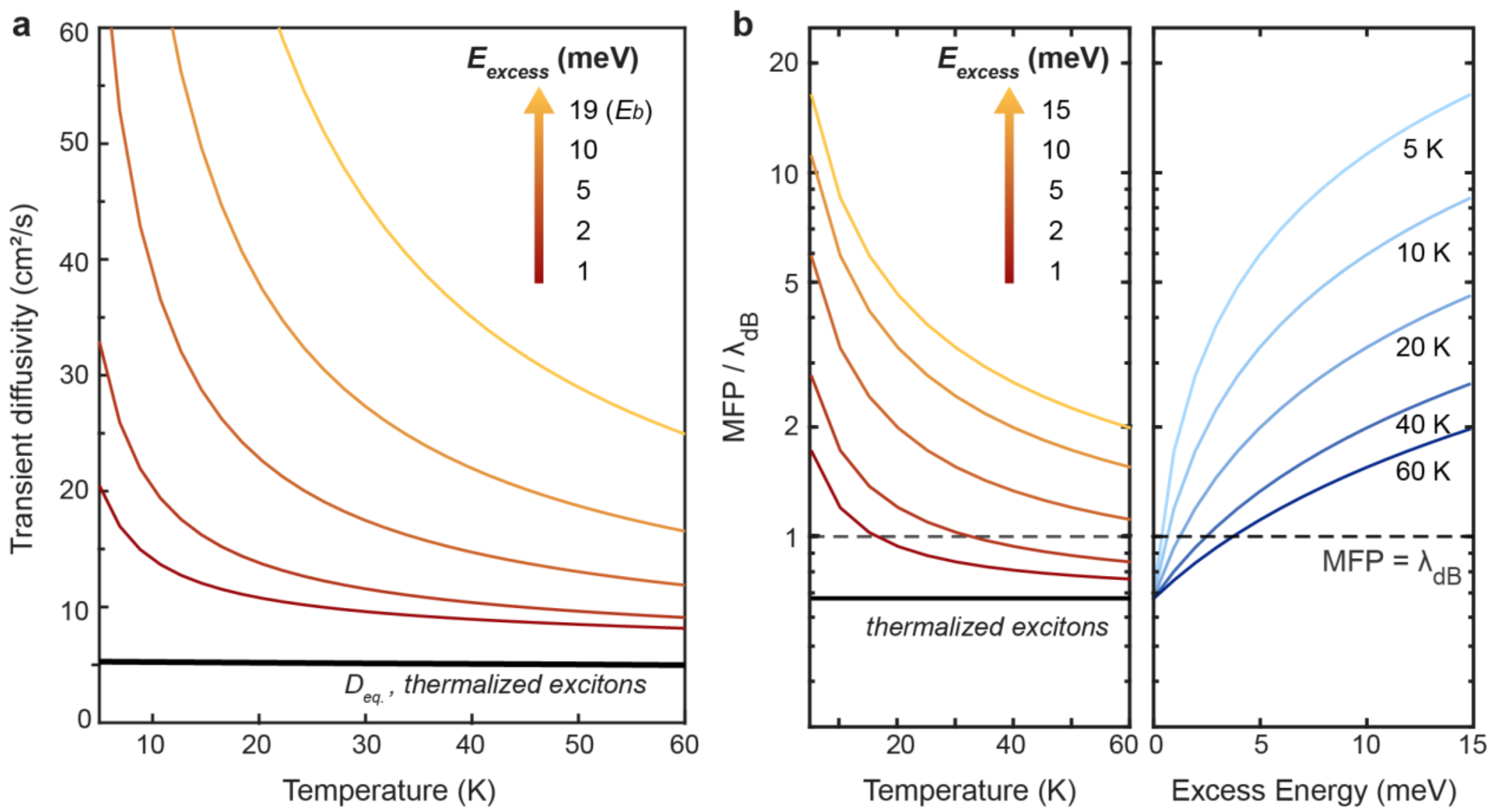

**Extended Data Fig. 7 | Excess kinetic energy enhances exciton transport. (a)** Predicted transient diffusivity for excitons with initial excess kinetic energy between 1 meV and $E_b$ (~19 meV). Hot-exciton diffusivity decreases with increasing temperature due to the temperature dependence of the scattering time. This behavior contrasts with the nearly temperature-independent diffusivity expected under equilibrium conditions (Fig. 3d). **(b)** Exciton mean free path (MFP) normalized by the thermal de Broglie wavelength ($\lambda_{\mathrm{dB}}$) as a function of temperature (left) and excess energy (right). The dashed line marks the Mott–Ioffe–Regel condition, $\mathrm{MFP} = \lambda_{\mathrm{dB}}$. Excess kinetic energy increases $\mathrm{MFP}/\lambda_{\mathrm{dB}}$, pushing exciton motion further into the band-like transport limit, whereas thermalized excitons remain close to the quantum localization threshold. For further discussion on the kinetic-theory calculations, see SI section 8.

# Methods

## Material growth and sample fabrication

*Vapor-phase epitaxy:* Epitaxial $CsPbBr_3$ single crystals were grown following a previously reported procedure.[33] The source material was prepared from cesium bromide (CsBr) and lead(II) bromide ($PbBr_2$) precursors, both of 99.999% purity (Sigma-Aldrich) mixed at a molar ratio of 2.6:1 ($CsBr:PbBr_2$). The mixture was placed in a lidded alumina boat and baked at 550 °C for 1 h under ultra-high-purity Ar gas flowing at 20 sccm. The off-stoichiometric precursor ratio compensates for differences in individual precursor evaporation rates and yields stoichiometric $CsPbBr_3$ crystals during vapor-phase growth.

Muscovite mica substrates (Electron Microscopy Sciences, V-0) were cut and cleaved in ambient air and cleaned by UV-ozone treatment immediately prior to growth. The mica substrates and the alumina boat containing the source powder were loaded into a custom CVD reactor consisting of a 1-inch-OD quartz tube housed in a horizontal two-zone furnace. Ultra-high-purity He was used as the carrier gas with flow rate and pressure of 100 sccm and 0.1 bar.

The source material (200 mg) was heated to 535 °C to melt, while the mica substrate was initially heated to 540 °C to remove surface contaminants and prevent premature nucleation. $CsPbBr_3$ growth was initiated by lowering the substrate temperature to 500 °C, establishing a temperature gradient along the substrate opposite to the carrier gas flow. The growth proceeded for 20 min and was abruptly terminated by increasing the chamber pressure to ~1 bar and removing the source boat from the hot zone without changing the substrate position. The furnace was then turned off and the reactor allowed to cool slowly to room temperature.

*Parylene-N deposition:* $CsPbBr_3$ films on mica were coated with parylene-N (thickness 0.2 - 1.1 μm) using a custom-built parylene deposition system. Details of the coating system and deposition procedure can be found at Ref [78]. Parylene forms chemically stable, conformal polymer films deposited at room temperature and provides electrically insulating encapsulation layers. In $CsPbBr_3$ FETs, parylene coating induces mild charge-transfer doping at the perovskite–polymer interface, producing a positive onset voltage and a noticeable dark conductivity in ungated devices.[33]

*Transmission electron microscopy (TEM)* characterization: Cross-sectional TEM specimens were prepared by focused ion beam milling. Initial trenching and lift-out were performed at room temperature using a FEI Helios Nanolab 600 DualBeam system with gallium ions, producing coarse micron-thick lamellae. Final thinning to an electron-transparent thickness of approximately 50–100 nm was carried out using a Zeiss Crossbeam FIB equipped with a cryo-cooled $N_2$ stage to reduce beam-induced damage and preserve the perovskite interface. HAADF-STEM imaging and EDS mapping were performed on an aberration-corrected Thermo Fisher Scientific Themis G3 equipped with Super-X EDS detectors, operated at 200 kV with beam currents of 80–150 pA.

**Electrical transport measurements**

Charge-carrier mobility was measured using field-effect transistor (FET) and gated Hall-effect methods. Hall measurements were performed in both FETs and uncoated/ungated films, yielding comparable mobilities.[33] The Hall voltage was measured using a Keithley 6514 electrometer in a magnetic field of $B = \pm 0.5$ T generated by an electromagnet (GMW, 5403 Dipole, 76 mm).Temperature-dependent mobility, $\mu(T)$, was measured between 50 – 320 K using a closed-cycle helium cryostat (Advanced Research Systems). Reversibility of $\mu(T)$ dependence between 200 and 320 K has been verified by repeatedly cycling the temperature in this range. Below ~ 100 K, $CsPbBr_3$ films on mica develop fine cracks, leading to gradual but irreversible device degradation. Hall mobilities in FETs were measured at gate voltages of $V_G = 0$ and -50 V, with no significant difference observed between these conditions. Four-probe FET mobilities were determined by fitting the linear region of the transfer characteristics over the gate-voltage range $V_G = -100 - 0$ V. Additional details of the electrical characterization are provided in Ref. [33]

**Optical transport measurements and spectroscopy:**

*Transient Photoluminescence Microscopy (TPLM):* Epitaxial $CsPbBr_3$ films were mounted on a piezoelectric stage (attocube ANC350) inside a closed-cycle helium optical cryostat (Montana Instruments Cryostation S100) and maintained under high vacuum (<0.1 mTorr). For temperature-dependent measurements, samples were cooled to 5 K and stabilized at the desired temperature using a controller (Lakeshore Model 335).TPLM measurements were performed using a home-built epifluorescence microscope (schematic in Ref. [ [75]]). Two optical excitation sources were used depending on the measurement conditions: (i) a Ti:sapphire laser operating at 76 MHz that was frequency-doubled to 405 nm (Coherent Mira 900, pulse width ≈150 fs), and (ii) a 405 nm pulsed laser diode operating at 5 MHz (PicoQuant LDH-P-C-405M, pulse width <100 ps). Unless otherwise stated, the excitation fluence at the sample surface was varied between 0.05-1 μJ $cm^{-2}$. The Ti:sapphire source was used for measurements below 60 K owing to its superior temporal resolution, while the pulsed diode was employed at higher temperatures.

The excitation beam was spatially filtered through a single-mode optical fiber and focused onto the sample to a near diffraction-limited spot using a cryostat-mounted objective (Zeiss EC Epiplan-Neofluar 100×, NA = 0.85). Photoluminescence was collected by the same objective in an epifluorescence configuration and filtered using a dichroic mirror (Semrock Di02-R405) and long-pass filter (Thorlabs FGL435M) to suppress scattered excitation light. The emission was relayed through a tube lens (Thorlabs TTL200-S8) and telescope optics (Thorlabs AC254-030-A and AC254-125-A), producing a total magnification of 495× at the detection plane.

Time-resolved detection was performed using a single-photon avalanche diode (Micro Photon Devices; timing resolution ≈50 ps; active area 50 μm × 50 μm) positioned at the magnified image plane. The detector was scanned along a single spatial dimension through the center of the photoluminescence profile, and time-correlated single-photon counting (TCSPC) histograms were acquired at each position. The spatiotemporal evolution of the photoluminescence profile was reconstructed from the resulting set of decay traces. For spectrally filtered transport measurements, a tunable notch filter (Semrock FL-445/16) mounted in a precision rotation stage was used to selectively suppress different portions of the emission spectrum.

*Carrier Diffusion Analysis:* In the low-excitation density regime, the spatiotemporal evolution of the photoexcited carrier population is described by the general diffusion equation (see also SI section 3.1):

$$\frac{\partial n(x,t)}{\partial t} = \nabla \cdot \left(D \nabla n(x,t)\right) - kn(x,t) \tag{1.1}$$

where $n(x,t)$ denotes the carrier density distribution as a function of position and time, $D$ is the effective carrier diffusivity, and $k$ is the first-order recombination rate. Over the short time window used to extract diffusivities (up to $\sim$ 1ns), both $D$ and $k$ are assumed to be time independent. Treating the laser pulse as an instantaneous excitation that generates an initial carrier distribution $n(\tilde{x},0)$, the solution to Eq. (1.1) can be written as:

$$n(x,t) = e^{-kt} \frac{1}{\sqrt{4\pi D t}} \int_{-\infty}^{\infty} n(\tilde{x},0) e^{-\frac{(x-\tilde{x})^2}{4Dt}} d\tilde{x} \tag{1.2}$$

which corresponds to a convolution of the initial distribution with the Gaussian Green's function of the diffusion equation. In many analyses the initial profile n(x,0) is assumed to be Gaussian, allowing diffusivity to be extracted directly from the temporal broadening of the photoluminescence (PL) profile. However, because the experimentally measured initial PL profiles deviate from a Gaussian shape, we employ a numerical convolution approach introduced by Grumstrup et al.[79] . In this method, each PL profile at t > 0 is fit by convolving the measured initial profile n(x,0) with a Gaussian kernel of time-dependent variance $\sigma^2$(t), without assuming a specific functional form for the excitation profile. The mean-squared displacement (MSD) is then obtained from the increase in the spatial variance relative to the initial profile:

$$\mathrm{MSD}\ (t) = \sigma^2(t) - \sigma^2(0) \tag{1.3}$$

At intermediate-to-high temperatures (~80-300 K), the MSD exhibits diffusive character and grows linearly over time. Therefore, the diffusivity along the analyzed direction is extracted from the slope of the MSD over the first nanosecond:

$$D = \frac{\Delta MSD}{2\Delta t} \quad (1.4)$$

At low temperatures, as the MSD becomes nonlinear, a single time-independent diffusivity does not adequately describe the transport. We therefore extract effective diffusivities from the local time-dependent MSD slope, $D(t) = \frac{dMSD}{2dt}$ (see Fig. S11) . The early-time diffusivity, $D_{\text{early}}$, is defined as the average effective diffusivity over a 40-ps window centered on the early-time maximum in $D(t)$ to reduce sensitivity to noise. The late-time diffusivity, $D_{\text{late}}$, is obtained from a linear fit to the MSD between 250-500 ps.

*Spectrally resolved PL dynamics and steady-state spectroscopy:* Spectrally resolved PL measurements were performed under the same optical excitation and sample configuration used for the TPLM experiments. After the tube lens, the emitted PL was coupled into a multimode optical fiber (Thorlabs BFL105LS02) and spectrally dispersed using a monochromator equipped with a 300 gr $mm^{-1}$ diffraction grating (Princeton Instruments SP2500). For time-resolved spectral measurements, the emission wavelength selected by the monochromator was detected with a single-photon avalanche diode (Micro Photon Devices; timing resolution ≈50 ps; active area 50 μm × 50 μm), and time-correlated single-photon counting (TCSPC) histograms were acquired sequentially at each wavelength. For steady-state measurements, the same optical pathway was used, with the monochromator output directed to a gated intensified CCD camera (Princeton Instruments PI-MAX4) to record time-integrated emission spectra.

*Temperature-dependent absorption spectroscopy:* Absorption spectra of epitaxial $CsPbBr_3$ crystals were measured in transmission geometry using a Cary 5000 UV–vis spectrophotometer. Samples were mounted in an optical cryostat (Janis ST-100) and cooled with liquid helium, enabling measurements between 4 and 300 K.

## Acknowledgments

Massachusetts Institute of Technology – Optical spectroscopy and transport measurements at MIT were supported by the US Department of Energy, Office of Science, Basic Energy Sciences under award number DE-SC0019345 (S.S., T.J.S, M.R., W.A.T). S.S acknowledges partial support from the Swiss National Science Foundation under Grant Number P500PN_202653. We also thank Dr. Mostapha Dakhchoune, Dr. Soroosh Sharifi-Asl, and Dr. Aubrey Penn for assistance with cross-sectional transmission electron microscopy sample preparation and imaging at the facilities of MIT.nano and the Harvard University Center for Nanoscale Systems.

Philipps Universität Marburg- Theoretical modelling was supported by the Deutsche Forschungsgemeinschaft (DFG) via the regular projects 504846924 and 542873285 (R.R, E.M).

Rutgers University- Epitaxial perovskite growth and sample preparation were supported by the US Department of Energy, Office of Basic Energy Sciences, Division of Materials Sciences and Engineering under award DE-SC0025401 (V.B, V.P). V.B and V.P additionally acknowledge the support received from RU Research Council for the growth equipment upgrades.

Supplementary Information for

# Nonequilibrium transport in epitaxial $CsPbBr_3$ single crystals

Seryio Saris[1], Roberto Rosati[2], Vladimir Bruevich[3], Thomas J. Sheehan[1], Maksim Román[1], Vitaly Podzorov[3], Ermin Malic[2], William A. Tisdale[1]

[1] Department of Chemical Engineering, Massachusetts Institute of Technology, Cambridge, MA, USA.

[2] Department of Physics, Philipps Universität Marburg, Marburg, Germany.

[3] Department of Physics and Astronomy, Rutgers University, Piscataway, NJ, USA.

## TABLE OF CONTENTS

## List of Figures

## List of Tables

## S1. Quasi-equilibrium modeling of excitons, free carriers and transport

### S1.1 Saha equilibrium considerations

At thermal equilibrium, we use the normalized Boltzmann distributions in bulk semiconductors to write the electron, hole and exciton distribution as:

$$f_q^a = V N_a f_q^{a\circ} \quad (1.1)$$

with $a = X, e, h$ respectively for excitons and optically excited electrons and holes. Here $f_q^{a\circ}$ is the normalized equilibrium distribution of species $a$ with (center-of-mass) momentum **q,** which is isotropic ($f_{\boldsymbol{q}}^{a\circ} \equiv f_q^{a\circ}$ with $q = |\boldsymbol{q}|$ ) and at low densities is provided by the normalized Boltzmann distribution $f_q^{a\circ} = \frac{e^{-\frac{\hbar^2 q^2}{2M_a}/k_BT}}{f_0^\circ}$, with $f_0^\circ = \frac{V}{2\pi^3}\left(\frac{M_a k_B T}{\hbar^2}\right)^{3/2}$. Here $V$ is the volume, $N_a = \frac{1}{V}\sum_{\boldsymbol{q}} f_{\boldsymbol{q}}^a$ are the corresponding spatial densities. $M_{e/h}$ and $M_X = M_e + M_h$ denote the corresponding carrier and exciton masses. We use the predicted temperature-dependent values,[1] with the room-temperature value of $M_h \approx 0.27m_0$, obtained via interpolation and in agreement with ARPES measurements of 0.20-0.26$m_0$, where $m_0$ is the free-electron mass.[2,3] In view of the ionization relation, at equilibrium these three spatial densities are related by the Saha equation[4,5]:

$$\frac{N_e N_h}{N_X} = \left(\frac{\pi k_B T}{(2\pi)^2\hbar^2}\frac{M_e M_h}{M_X}\right)^{\frac{3}{2}} e^{-\frac{E_b}{k_B T}} \quad (1.2a)$$

$$N_e = N_h + n_e^d \quad (1.2b)$$

$$N = N_h + N_X \quad (1.2c)$$

where $N$ is the total density of optically excited pairs, while $n_e^d$ indicates the presence of residual electron doping (without loss of generality, similar results would be obtained for hole doping). The relative ratio among the three species is crucially determined by the lattice temperature $T$ and the exciton binding energy $E_b$.[6] Here $m = \frac{M_e M_h}{M_e + M_h}$ is the reduced mass while $\epsilon$ is the effective background dielectric constant, for which we take an effective value $\epsilon \approx 8$ to recover the binding energy of ~19 meV found at cryogenic temperatures.[1,7] Note that Eqs.1.2 hold provided that the low-density limit is still valid (otherwise the Boltzmann distribution should be replaced by the corresponding quantum ones).

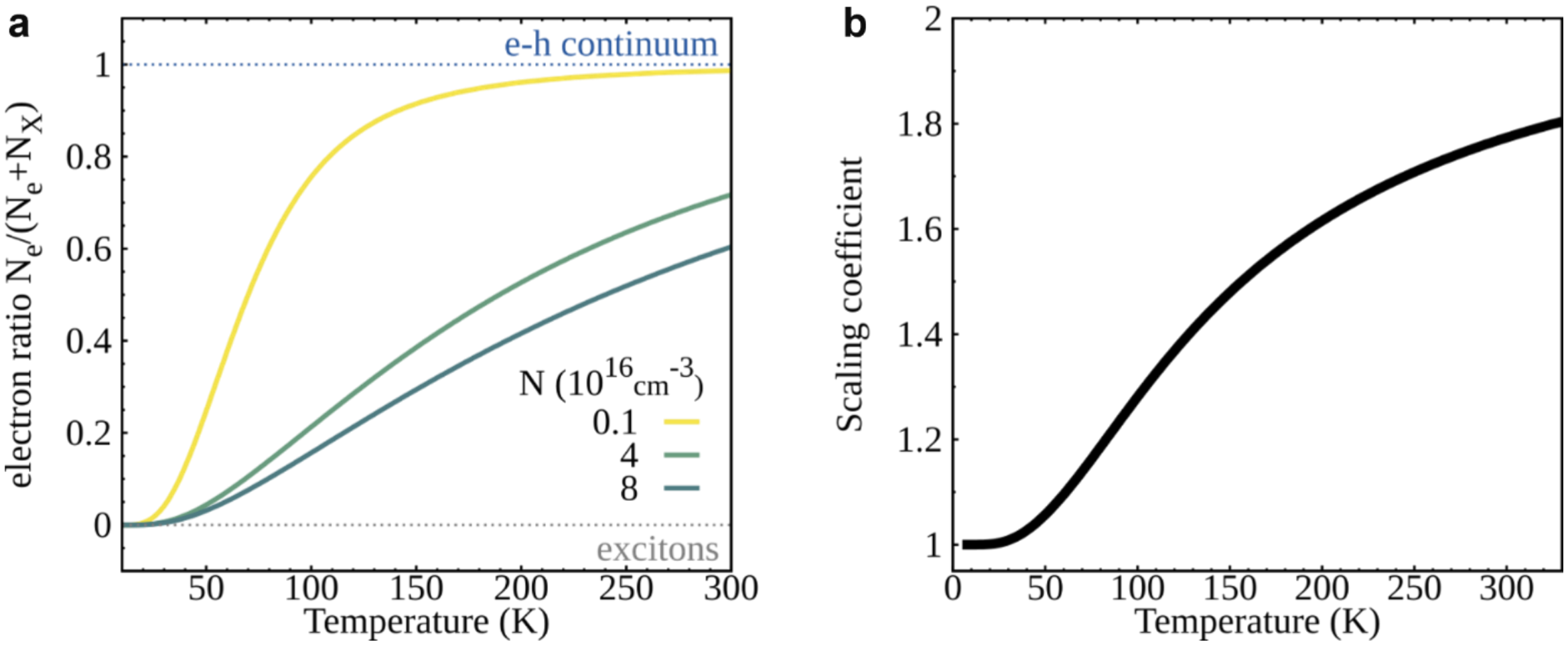

**Figure S1 | Free carrier–exciton equilibrium and recombination scaling.** (a) Free-carrier fraction extracted from the Saha equilibrium for three different excitation densities. Higher temperatures trigger the fission of low-energy excitons into the energetically higher electron-hole pairs, while increasing pump power favors the fusion of electrons and holes into excitons by decreasing their average spatial separation. (b) Corresponding fluence-scaling exponents *k* expected from the relative free-carrier and excitonic populations. Both quantities show a crossover from a predominantly free carrier population at higher temperatures, $k \approx 2$ associated with bimolecular recombination, to an exciton-rich regime at lower temperatures where monomolecular recombination dominates ($k \approx 1$).

---

In Fig. S1 we investigate how the fraction of electron-hole pairs relative to the optically-excited quasiparticles is affected by excitation density as well as temperature. At low temperatures, the excitons dominate, as the thermal energy is smaller than the binding energy, resulting in a negligible fission of excitons into the high-energy continuum. At room temperature, this balance shifts toward the continuum: the larger thermal energy promotes exciton dissociation and populates weakly bound electron–hole pairs, whose much larger density of states outweighs the single 1s exciton level. Although this qualitative trend is preserved by excitation density, higher pump power results in a delayed transition from exciton-dominated to a continuum-dominated system. This stems from Saha's mass action law, Eqs. 1.2a, and can be understood as enhanced exciton formation at higher excitation densities, where the average electron–hole separation is smaller.

The temperature- and density-dependent variations in the excitonic and continuum fraction can be experimentally accessed via the scaling of energy-integrated emission, $I \propto N^k$, with the excitation density *N*. In exciton-dominated systems, the scaling coefficient approaches $k \approx 1$, as the exciton population is proportional to the number of absorbed photons. In contrast, in free-carrier-dominated systems with negligible doping, the recombination rate scales with the product of the electron and hole densities, giving a coefficient approaching $k \approx 2$. Both limiting behaviors are observed in our measurements, with $k \approx 1$ in the low-temperature excitonic regime and $k \approx 2$ at high temperatures, where thermal dissociation favors unbound electron–hole pairs (Fig. 2, main text).

### S1.2 Modeling of free carrier, exciton and total diffusivities

In the limit of locally quasi-thermalized distributions, the electron and hole diffusion coefficients, $D_{e/h}$ read[8,9]:

$$D_{e/h} = \frac{1}{3} \sum\nolimits_{\boldsymbol{q}} \tau_{\boldsymbol{q}}^{e/h} \left( \frac{\hbar q}{m_{e/h}} \right)^2 f_{\boldsymbol{q}}^{e/h,\circ} \tag{1.3}$$

where $\tau_{\boldsymbol{q}}^{e/h} \equiv \tau_{q}^{e/h} = \hbar/\Gamma_q^{e/h}$ is the isotropic state-dependent scattering time obtained from the corresponding scattering rate $\Gamma_q^{e/h}$. These rates $\Gamma_q^{e/h} = \Gamma_q^{ac.,e/h} + \Gamma_q^{Fr.,e/h}$ are obtained by summing the contribution due to scattering with acoustic and optical phonons via Fröhlich interaction, respectively, $\Gamma_q^{ac.,e/h}$ and $\Gamma_q^{Fr.,e/h}$. The state-dependent scattering rates due to Fröhlich interaction are obtained via the conventional Markov approximation $\Gamma_{\boldsymbol{q}}^{Fr.,e/h} \equiv \Gamma_q^{Fr.,\frac{e}{h}} = 2\pi\hbar \sum_{\mathbf{q}',\pm} \left| g_{\mathbf{q}-\mathbf{q}'} \right|^2 n_{\mathrm{LO}} \delta\left( \frac{\hbar^2 q^2}{2M_{e/h}} \mp E_{\mathrm{LO}} \right)$, where $n_{LO}$ is the occupation of LO phonon with energy $E_{LO} \approx 18$ meV[1] while $g_{\boldsymbol{q}}$ is the coefficient of the electron-phonon Fröhlich Hamiltonian:[6]

$$g_{\mathbf{q}} = \frac{e}{q} \left[ \frac{1}{4\pi\epsilon_0 V} 2\pi E_{\mathrm{LO}} (\epsilon_\infty^{-1} - \epsilon_0^{-1}) \right]^{\frac{1}{2}} \tag{1.4}$$

where $\epsilon_0 = 16$ and $\epsilon_\infty = 4.5$ are, respectively, the low- and high-frequency dielectric constants.[1] The rate due to scattering with acoustic phonons $\Gamma_q^{ac.,e/h} \equiv \Gamma^{ac.} = c_{ac} T$ is assumed to be only weakly state-dependent and taken from the experimental linewidth with $c_{ac} = 50\mu eV/K$ (see section 4.2). The resulting diffusion of the electron-hole plasma $D_{e/h}$ is obtained, assuming pure ambipolar regime under nondegenerate conditions as in Ref [8]:

$$D_{e/h}^{-1} = \frac{1}{2} \left( D_e^{-1} + D_h^{-1} \right) \tag{1.5}$$

Analogously the excitonic diffusion coefficient $D_X$ reads:

$$D_X = \frac{1}{3} \tau_X \sum\nolimits_{\boldsymbol{q}} \left( \frac{\hbar q}{M_X} \right)^2 f_{\boldsymbol{q}}^{X,\circ} \tag{1.6}$$

Here we have assumed that the Fröhlich interaction has a negligible impact on excitons (due to cancellation between electron and hole contributions),[10,11] resulting in the state-independent scattering time dominated by acoustic phonons, $\tau_X = \hbar/\Gamma^{ac}$. Finally, the overall diffusion coefficient $D$ is obtained from the weighted average of electron/hole and exciton diffusion coefficient as in Ref [5]:

$$D = \frac{N_X}{N_X + N_{eh}} D_X + \frac{N_{eh}}{N_X + N_{eh}} D_{eh} \tag{1.7}$$

where $N_{eh}$ is the density of optically excited carrier, for which $N_{eh} \equiv N_h \approx N_e$ holds for weak doping ($n_e^d \ll N_h$). Both densities $N_{eh}$ and $N_X$ depend on temperature and excitation density according to the Saha equation, cf. Eq. 1.2a and Fig. S1.

In Fig. S2a we look at the resulting temperature-dependent equilibrium mobility $\mu = \frac{D}{k_B T}$ for different pump powers at intermediate and high temperatures. For small excitation densities the system is dominated by the unbound electron-hole pairs, Fig. S1, and the overall diffusion $D$ coincides with the continuum ambipolar diffusivity $D_{eh}$ (yellow and dotted-blue line, respectively). Increasing the excitation densities favors the formation of excitons via a decreased electron-hole separation, cf. Fig. S1, and the total diffusion becomes increasingly dominated by the intrinsic diffusion coefficient of excitons upon cooling (dark green and light gray line, respectively). Despite having larger masses, at high temperatures the excitons show a larger diffusivity than electron-hole pairs thanks to the negligible scattering with optical phonons via Fröhlich interaction, as this leads to cancelling effects between the electron and the hole constituting the exciton.

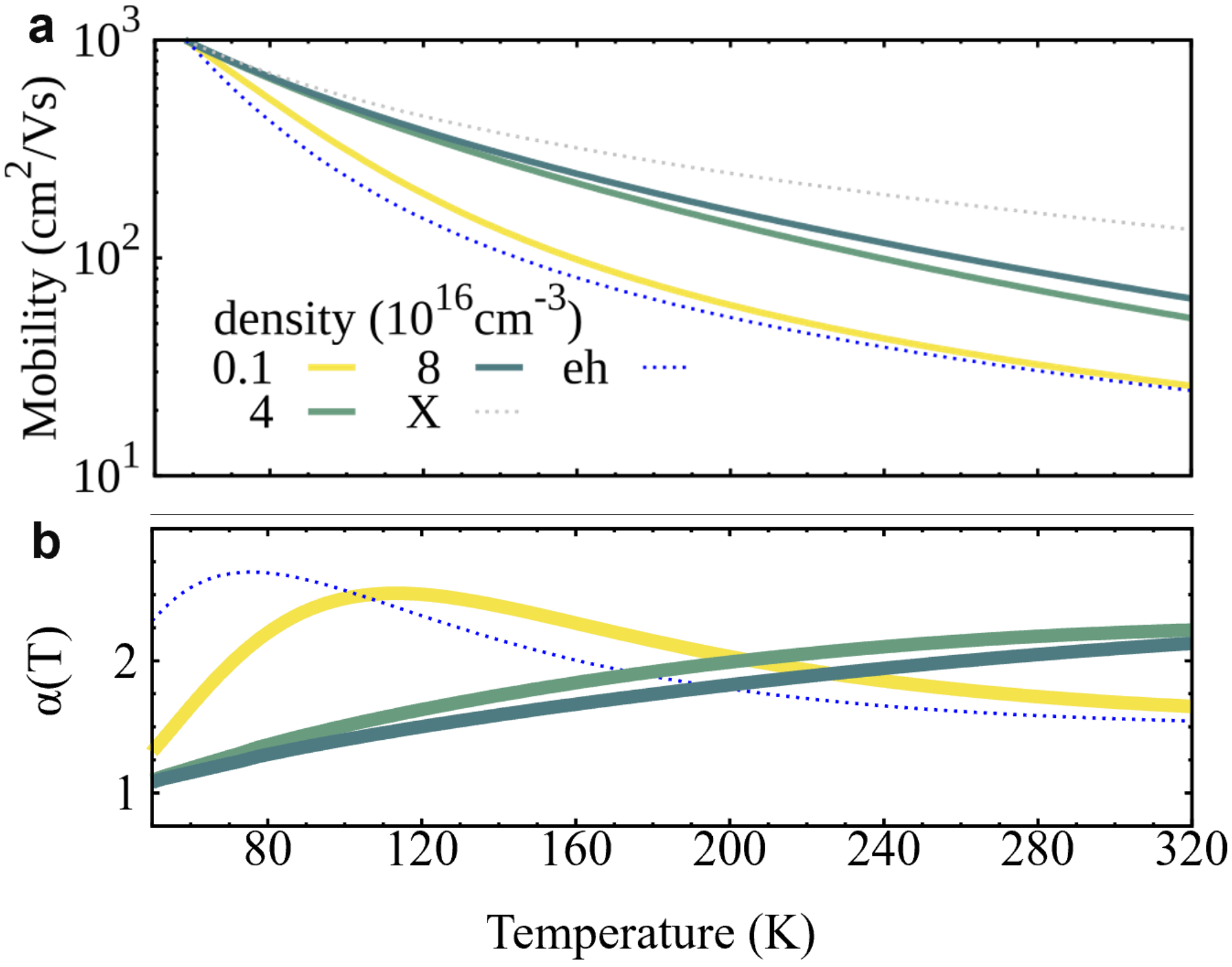

**Figure S2 | Temperature-dependent mobility at thermal equilibrium. (a)** Steady-state mobility $\mu$ calculations as a function of temperature $T$. Solid lines show the total equilibrium mobility obtained by the Saha-weighting of the exciton and electron-hole continuum contributions at three photoexcitation densities. The limiting cases of a purely excitonic and purely electron-hole continuum are shown as dotted gray and blue lines, respectively. The overall occupation is a mixture of exciton and continuum, leading to mobilities lying in between the one corresponding to each of the two species. **(b)** Scaling exponent $\alpha = -\frac{T}{\mu}\partial_T \mu$ as a function of temperature $T$, with $\alpha$ such that the mobility scales as $\mu = \mu_0 T^{-\alpha}$ with $T$.

In the gedanken case of temperature-independent diffusion coefficients, the mobility scales as $\mu = \mu_0 T_0/T$, with $\mu_0 = D/k_B T_0$ being the mobility at a reference temperature $T_0$. However, in reality the diffusion coefficients $D$ vary with temperature due to i) thermal activation of the scattering with different phonon modes and ii) the temperature-dependent ratio between exciton and electron-hole continuum. The non-trivial variation of diffusion with temperature leads to mobilities behaving as $\mu \propto T^{-\alpha}$, with the exponent $\alpha = -\frac{T}{\mu}\partial_T \mu$ differing in general from 1.

In Fig. S2b, we look at the corresponding exponent $\alpha$, where we observe different behavior with temperature and excitation power. At very low temperature, we recover $\alpha = 1$, as the system is dominated by excitons whose diffusion constant becomes temperature independent at low temperatures, unless quantum effects take over (see Fig. S13 and section S8 for further discussion on the low-temperature regime).[12,13] At room temperature, we find different behavior according to the excitation power. In this low-density regime the system is dominated by electron-hole pairs, whose diffusivity presents a residual decrease with increasing temperatures. This happens because higher thermal energies result in a larger occupation of high-energy electrons and holes, which have higher scattering rates because they are energetically allowed to emit optical phonons. Overall at small excitation densities we find at room temperature an exponent $\alpha \approx 1.7$, in excellent agreement with our TPLM measurements, as well as recent independent experiments.[14] Larger excitation densities increase the amount of excitons, which have a higher mobility because they are weakly affected by Fröhlich interaction. In this regime increasing temperatures lead to the dissociation of excitons into less-mobile electron-hole pairs, resulting in a faster decrease of mobility with temperature and, hence, in larger exponents $\alpha$ (green lines).

The situation becomes more involved at intermediate temperatures, where the mobility is affected by i) the activation of Fröhlich interaction for electron-hole pairs and ii) the competition between exciton and continuum, whose relative weight depends on excitation power. At small excitation powers we find exponents $\alpha \approx 2.5$ at $T \approx$100 K. Such a large value decreases at higher excitation powers, because in this case most of the electron-hole pairs have already bound into excitons, which furthermore are weakly affected by Fröhlich interaction. Consequently, at higher excitation densities the exponent approaches the cryogenic value of $\alpha \approx 1$ already at intermediate temperatures.

### S1.3 Microscopic PL model including excitonic and continuum contributions

The expected photoluminescence intensity can be calculated by:

$$I(E - E_{bg}) = \frac{g}{V}\left|\sum_{\boldsymbol{k}} \phi_{1s}(\boldsymbol{k})\right|^2 f^X_{q=0} \frac{\Gamma_X}{(E + E_b)^2 + \Gamma_X^2} + g\sum_{\boldsymbol{q}} \frac{1}{V} f^e_q f^h_{-q} \frac{\Gamma_q}{\left(E - \frac{\hbar^2}{2m}q^2\right)^2 + \Gamma_q^2} \tag{1.8}$$

where we used that the oscillator strength is proportional to the effective optical dipole $g_q \equiv g$ times the probability of finding an electron and hole in the same position. For the 1s exciton this reads $|\phi_{1s}(r=0)|^2 = \frac{1}{V}|\sum_k \phi_{1s}(k)|^2$, with $\phi_{1s}(r)$ being the wave-function in a relative position. For this we use $\phi_{1s}(r) = e^{-\frac{r}{r_0}}/\sqrt{\pi r_o^3}$,[6] and $\phi_{1s}(k) = c_{1s}(1 + k^2 r_0^2)^{-2}$ with normalization constant $c_{1s} = 1/\sqrt{\sum_{\mathbf{k}} |(1 + k^2 r_0^2)^{-2}|^2}$ . For electron-hole pairs such a probability gives $1/V$, as their wavefunctions are fully delocalized in the relative positions (Kronecker delta in momentum, after neglecting the exciton correction to their orthogonal plane waves[15]). Finally, $E_b$ and $\mu$ are the exciton binding energy and relative mass, while $\Gamma_X = \Gamma_0 + \Gamma^{ac}/2$ and $\Gamma_q = \Gamma_0+(\Gamma_q^e + \Gamma_q^h)/2$ provide the dephasing rate for exciton and electron-hole pair with corresponding momentum $q$. These are provided by the corresponding scattering-induced dephasing rates plus the temperature independent contribution $\Gamma_0 \approx 9$ meV as per experimental linewidth at $T = 0$.

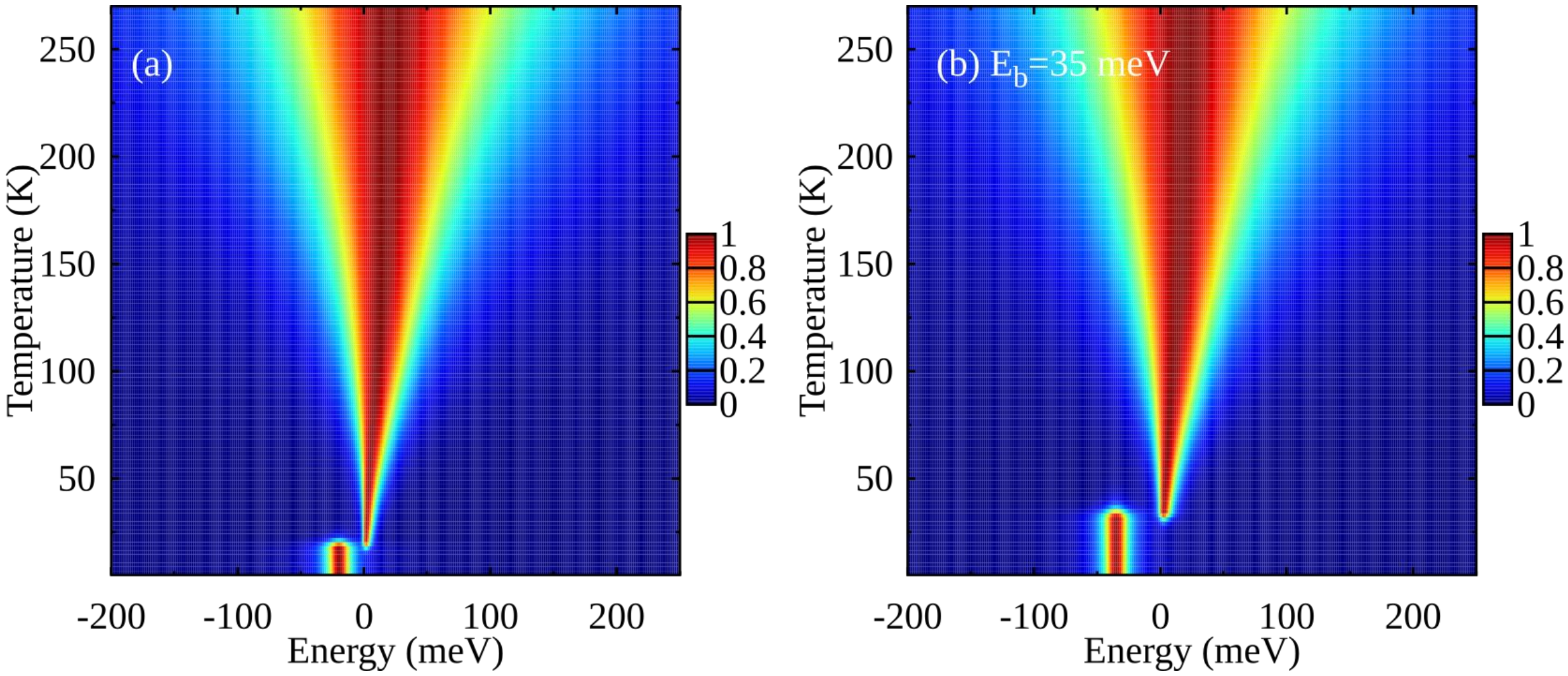

**Figure S3 | Modeling photoluminescence spectra.** Simulated PL spectra using the microscopic Elliott-type model, showing the redistribution of oscillator strength into the bound exciton state at low temperature. Panels compare spectra calculated using two different exciton binding energies: **(a)** the temperature-dependent polaronic binding energy used in our transport model, which accounts for Fröhlich-coupling-induced modifications of dielectric screening and carrier mass **(b)** the temperature-independent binding energy of 35 meV, representative of values commonly extracted for $CsPbBr_3$ using simplified hydrogenic exciton models.[34]

The first part of Eq. 1.8 coincides with the excitonic Elliott formula for the photoluminescence,[16] where we have included only the 1s bright exciton states in view of the relatively small binding energy. The second term of Eq. 1.8 provides the emission due to the continuum, which is analogous to the formula obtained in Ref. [17] here after neglecting Coulomb effects. Importantly, these would lead to strong plasma contributions also at the excitonic spectral peak,[18,19] which become particularly important for smaller temperatures and linewidths.[18,20] The omission of these effects could overestimate the signal spectrally above the band-gap, in comparison to the one at exciton resonance and in particular at low temperatures. In Fig. S3 we compare the resulting photoluminescence emission with the one evaluated in the case of a larger temperature-independent binding energy of 35 meV. The occupations are evaluated at equilibrium, i.e. without accounting for transient effects. At low temperatures the signal is dominated by the low-energy exciton feature, while at higher energies we see the appearance of the energetically higher peak emitted from the continuum. Such a transition takes place at higher temperatures in the presence of larger binding energies.

## S2. Sample morphology

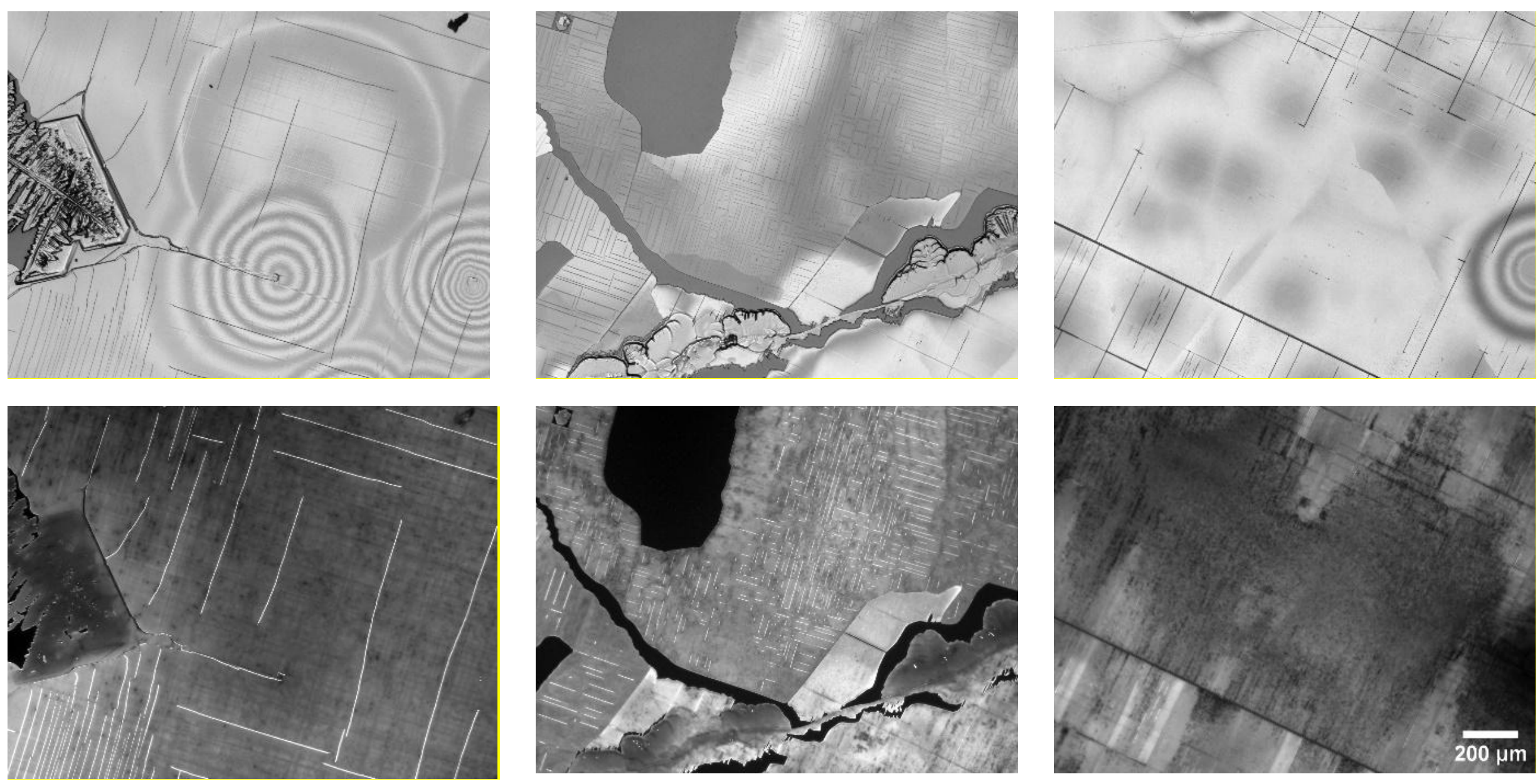

**Figure S4 | Optical microscopy images of sample surface** (**top row**) Bright field (BF) images, **(bottom row)** corresponding widefield photoluminescence (PL) images of epitaxial $CsPbBr_3$ films. Some grains contain cracks formed in the film during the cooling stage of the growth due to a mismatch in the coefficients of thermal expansion between the substrate and the crystalline film. These cracks appear as dark lines in BF and bright lines in PL images. Images were acquired with a Zeiss photoluminescence Axioplan 2 microscope.

## S3. Transport dynamics at the quasi-equilibrium regime

### S3.1 Note on the use of first-order diffusion equation

In transient photoluminescence microscopy experiments (TPLM), the spatial broadening of the PL profile is governed by the competition between diffusive spreading and density-dependent recombination. The relative importance of these two processes is captured by the dimensionless ratio $D/(\Delta x^2 B n_0)$, where $D$ is the diffusivity, $\Delta x$ the initial spatial width, $B$ the bimolecular recombination coefficient, and $n_0$ the peak carrier density.[21] When this ratio is much greater than unity, diffusion dominates and the spatial dynamics are well described by a first-order diffusion equation (see Methods Eq. 1.1), independent of the recombination kinetics.

For bulk $CsPbBr_3$ perovskites, reported diffusivities of $D = 0.5$–$1.5$ cm$^2$/s and bimolecular coefficients $B \sim 10^{-10}$–$10^{-11}$ cm$^3$/s, [22–24] combined with the carrier densities used here ($\leq 10^{16}$

cm$^{-3}$), place our measurements in the diffusion-dominated regime. This remains the case at intermediate and room temperatures, where the spectrally integrated emission exhibits superlinear fluence scaling ($k \approx 1.2$–$2$; Fig. 2c in main text) reflecting bimolecular recombination contributions to the recombination kinetics—contributions that do not appreciably broaden the spatial profile evolution from which diffusivities are extracted. This distinction has been validated in previous TPLM studies of bulk perovskites, where diffusivities obtained from first-order treatments agree quantitatively with full three-dimensional diffusion–recombination modeling.[21,25]

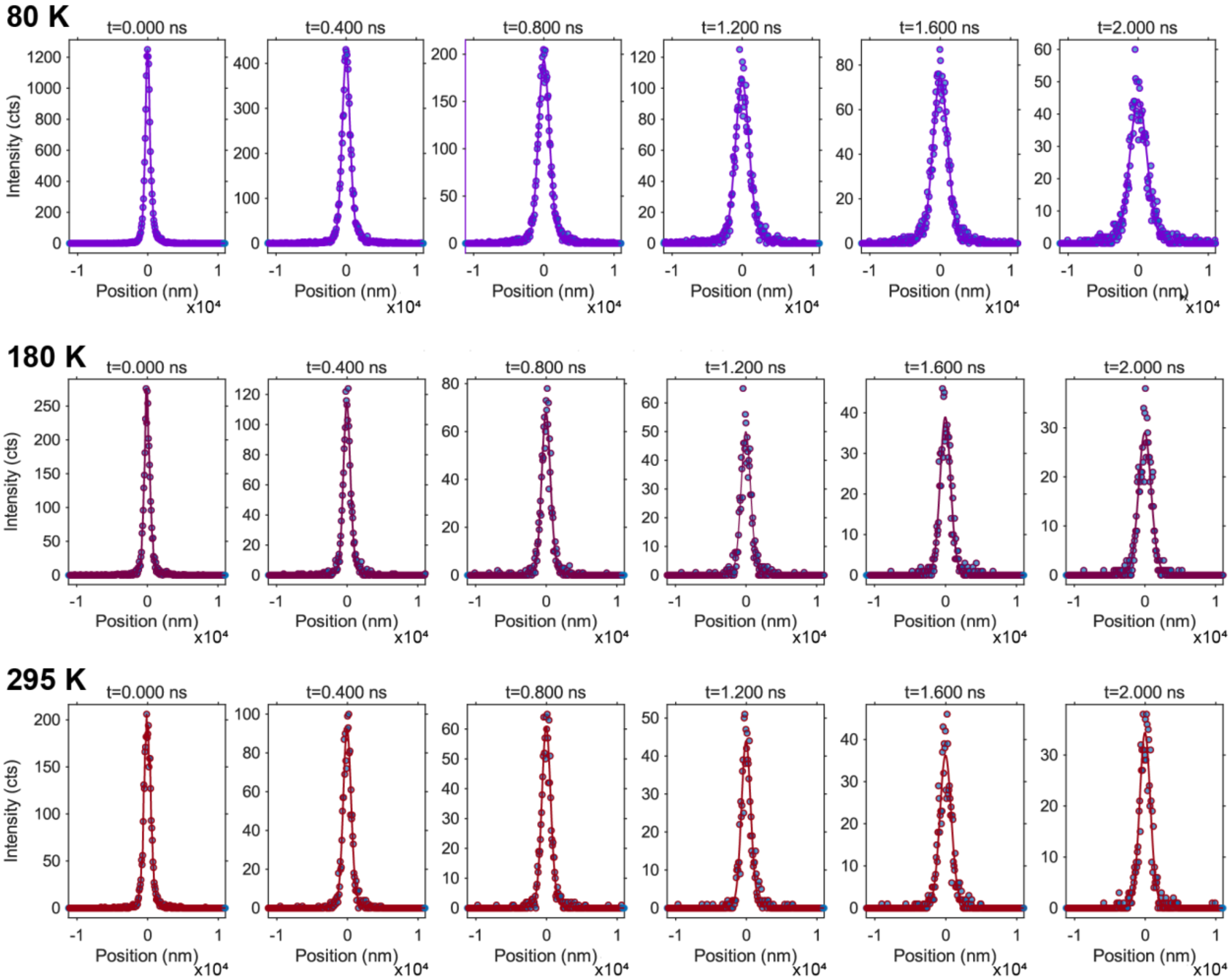

**Figure S5 | Temporal evolution of spatial PL profiles** at selected delay times for three temperatures. Symbols show the experimental PL intensity as a function of position. Solid lines are fits obtained by convolving the initial spatial profile at $t = 0$ with a Gaussian kernel, capturing the temporal broadening of the distribution. The spatial width increases with delay time as the photoexcited population expands laterally, with enhanced broadening observed at lower temperatures.

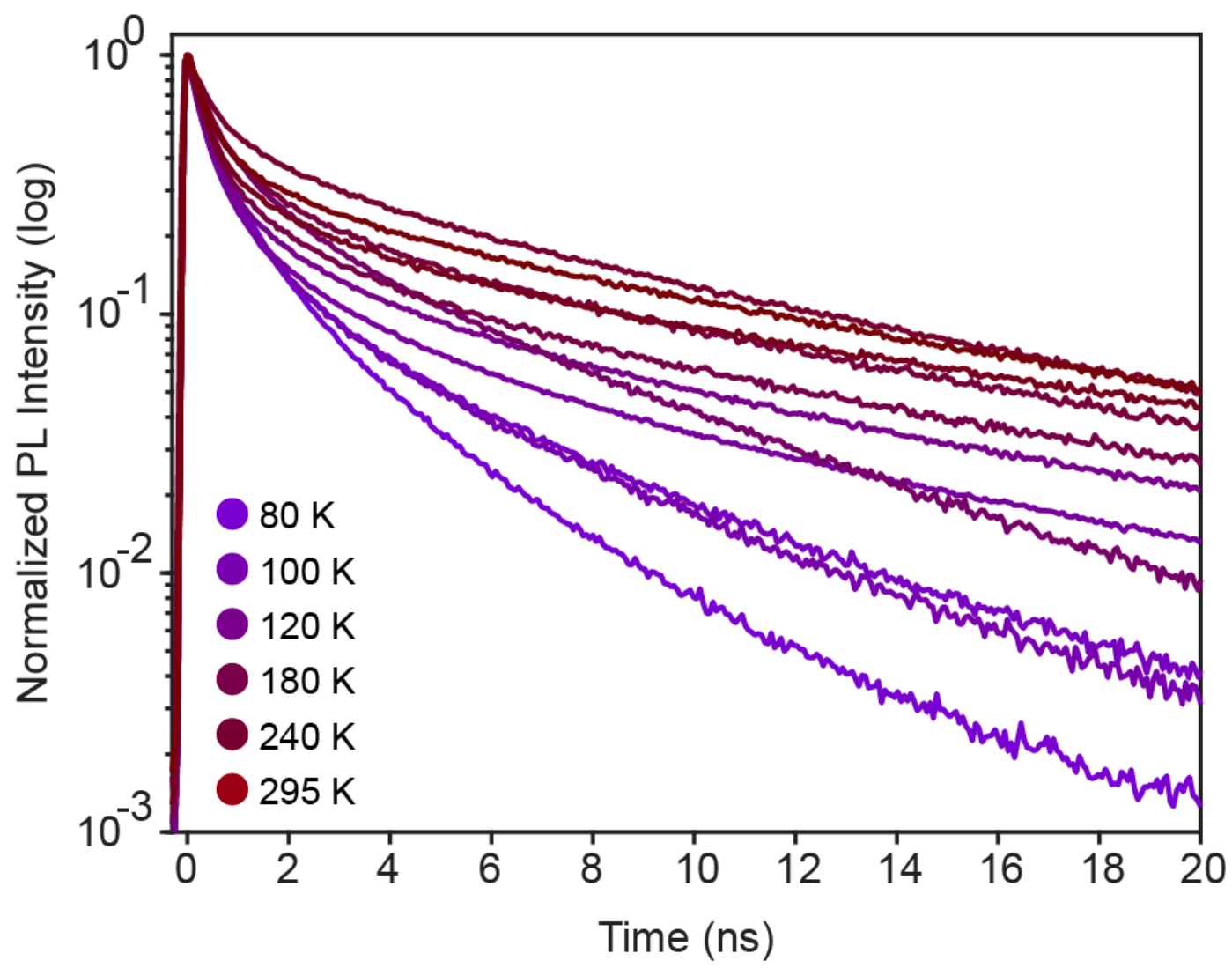

**Figure S6 | Extended dataset of spatially integrated PL dynamics at intermediate-high temperatures.** Decays measured between 80K- 295K (color-coded) on a logarithmic intensity scale. All traces are normalized to their peak intensity. The prompt decay dynamics accelerate with decreasing temperature, as reflected by the apparent $1/e$ lifetimes shown in the inset of Fig. 1d. The decays are non-exponential, with an initial rapid drop followed by pronounced long-time tails, a behavior widely observed in lead-halide perovskites and commonly associated with power-law-like recombination dynamics.[33] Accordingly, the 1/e times are used here as comparative descriptors of the prompt dynamics rather than as single-valued recombination lifetimes.

## S4. Spectroscopic signatures of the excitonic crossover

### S4.1 Enhanced optical quality via parylene encapsulation

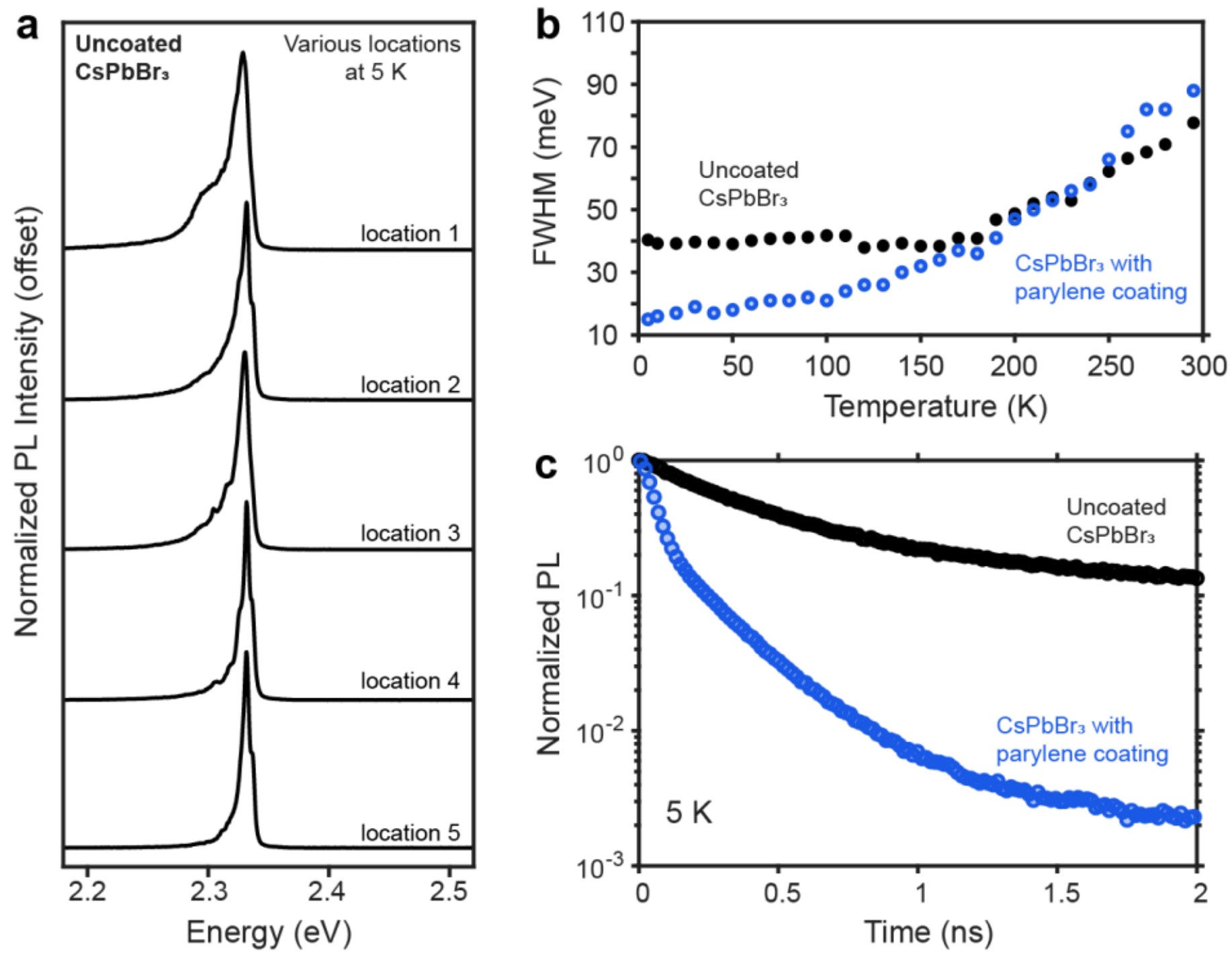

**Figure S7 | Parylene encapsulation improves optical quality of $CsPbBr_3$ (a)** Stacked, normalized PL spectra recorded at different spatial locations of an uncoated $CsPbBr_3$ crystal at 5 K show pronounced spectral heterogeneity, including variation in linewidth and multiple broad emissive features. Similar low-temperature multipeak emission has previously been attributed to trapped excitons and shallow disorder-induced band-tails arising from native defects, impurities and structural disorder.[26] **(b)** Temperature-dependent linewidths further show that uncoated crystals remain substantially broader at low temperatures, whereas parylene encapsulated crystal exhibit narrower emission (FWHM <15 meV) **(c)** Time-resolved PL at 5 K reveals substantially longer decay lifetimes for uncoated crystals relative to the parylene-coated sample, consistent with trap-related emission. Collectively, these observations suggest that unencapsulated samples exhibit spectral inhomogeneity and broader distributions of emissive states, which can partially obscure intrinsic low-temperature excitonic behavior that becomes more clearly resolved following surface passivation.

### S4.2 Analysis of temperature-dependent PL broadening

The temperature dependence of the PL linewidth shown in Extended Data Fig. 4d (main text) is analyzed using a phenomenological broadening model commonly employed to separate distinct phonon-scattering contributions in semiconductor emission spectra. The extracted full width at half maximum (FWHM), $\Gamma(T)$, is described as the sum of residual inhomogeneous broadening and acoustic- and optical-phonon contributions:

$$\Gamma(T) = \Gamma_0 + \gamma_{\mathrm{ac}} T + \frac{\gamma_{\mathrm{O}}}{\exp\left(E_{\mathrm{O}}/k_B T\right) - 1}. \quad (4.1)$$

Here, $\Gamma_0$ is the residual inhomogeneous linewidth, $\gamma_{\mathrm{ac}}$ is the acoustic phonon contribution, and $\gamma_{\mathrm{O}}$ is the coupling to optical phonons with characteristic energy $E_{\mathrm{O}}$. The acoustic contribution is treated phenomenologically as a linear-in-$T$ term, whereas optical phonon broadening follows Bose–Einstein statistics and exhibits nonlinear temperature dependence. Similar phenomenological analyses have been widely used to disentangle competing phonon-scattering channels in hybrid perovskites.[27–29]

The strong temperature-dependent increase in linewidth in Extended Data Fig. 4d is characteristic of efficient phonon-assisted broadening in bulk perovskites. The nonlinear increase at elevated temperatures is captured by the optical phonon term, which dominates the overall linewidth evolution over much of the measured temperature range. At lower temperatures, where the optical phonon population is strongly reduced, the residual linear linewidth increase provides an empirical estimate of acoustic-phonon broadening. From this low-temperature slope, we extract $\gamma_{\mathrm{ac}} = 50\ \mu\mathrm{eV/K}$, comparable to values reported for bulk lead halide perovskites. This acoustic broadening coefficient is used in Section S1 to constrain the acoustic-phonon scattering rate in our microscopic transport model.

### S4.3 Fluence-dependent PL spectra

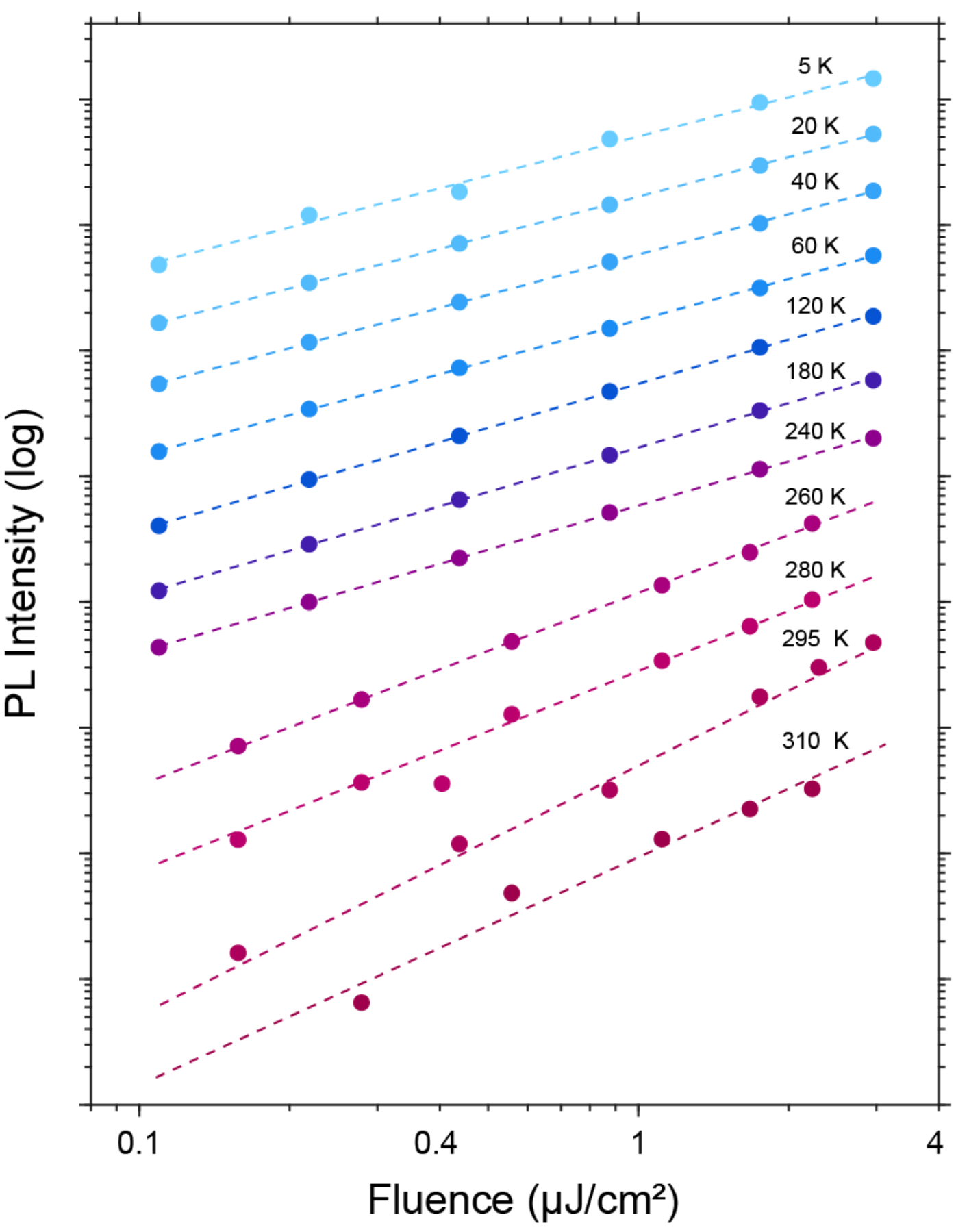

**Figure S8 | Fluence scaling range and coefficients**. Log–log plots of the total integrated PL intensity as a function of excitation fluence measured between 5–310 K. Dashed lines indicate linear fits in log–log space of the form $I \propto F^k$, from which the scaling exponent $k$ is extracted and summarized in Fig. 2c of the main text. The excitation conditions correspond to estimated average carrier densities spanning approximately $4 \times 10^{15}$to $2 \times 10^{17}\ \mathrm{cm}^{-3}$. The scaling exponent evolves from $k \approx 2$ near room temperature to $k \approx 1$ below ~60 K, consistent with a crossover from free carrier to excitonic recombination. The approximately linear power-law behavior observed across this fluence range confirms that the dominant recombination pathway is preserved over the probed excitation densities. Thus, the measurements are performed below the regime where strong many-body interactions, state filling, or higher-order recombination processes significantly modify the PL response.

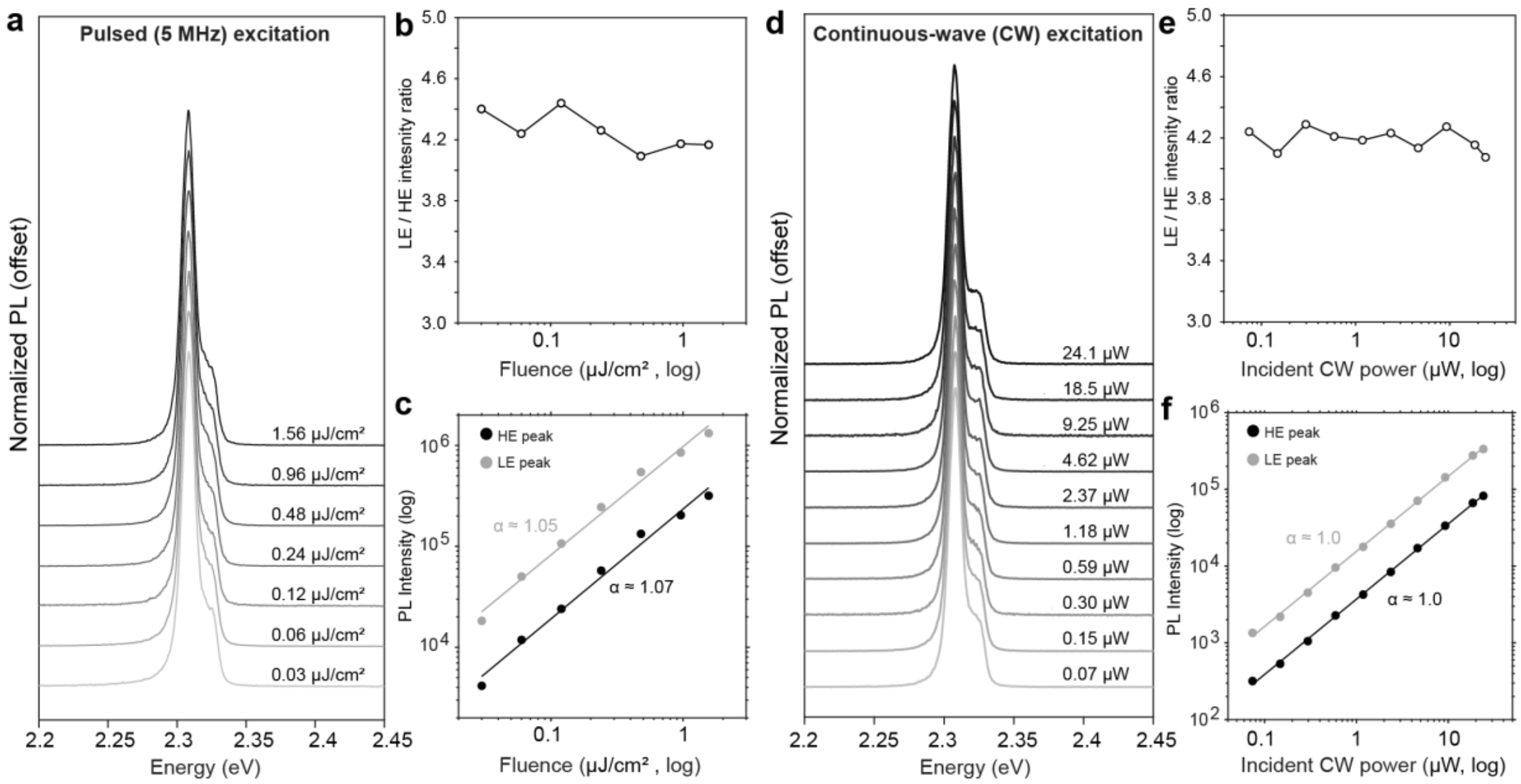

**Figure S9 | Fluence-dependent PL spectra at 5 K. (a)** Stacked PL spectra recorded under nonresonant p*ulsed excitation* at 5 MHz as a function of excitation fluence. **(b)** Fluence dependence of the intensity ratio between the low-energy (LE) and high-energy (HE) emission peaks. **(c)** Log–log plot of the LE and HE peak intensities versus excitation fluence. **(d–f)** Corresponding measurements performed under nonresonant *continuous-wave* (CW) excitation. Across both excitation modalities, the LE/HE ratio remains nearly constant, and both peaks exhibit comparable linear scaling, supporting an excitonic origin of the two emissive components and ruling out excitation-density effects such as trap saturation, state filling, or other higher-order recombination processes as their primary origin.

## S5. Transport dynamics at the nonequilibrium regime

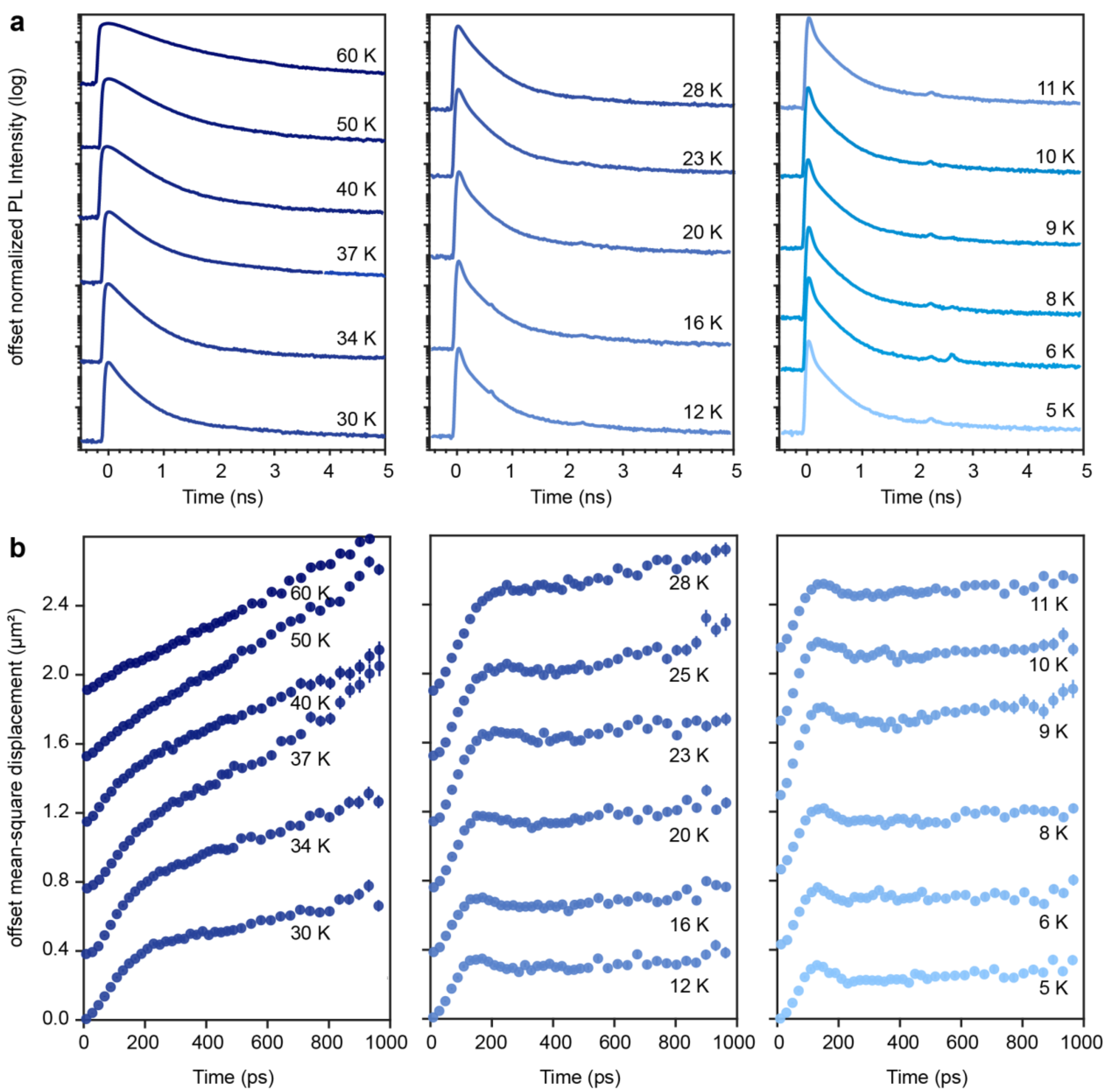

**Figure S10 | Extended low-temperature dataset of spectrally integrated PL dynamics and MSD evolution. (a)** spatially integrated PL decay dynamics measured from 5 to 60 K. Upon cooling, the PL response develops a fast sub-100 ps relaxation component. **(b)** Corresponding mean square displacement (MSD) dynamics across the same temperature range. Below ~40 K, the emergence of this fast PL component occurs together with a rapid initial MSD rise, followed by nonlinear MSD evolution and an apparent late-time contraction.

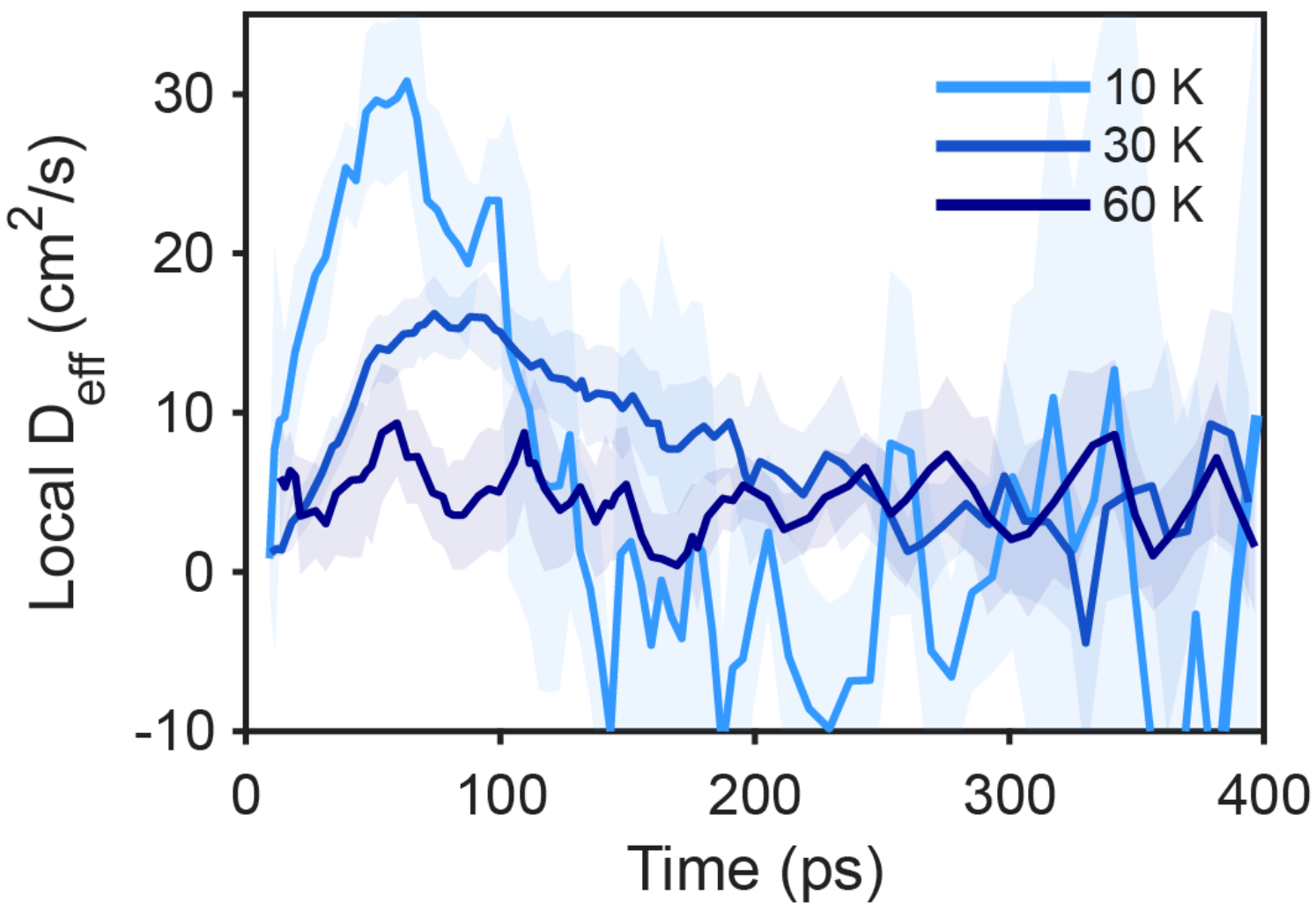

**Figure S11 | Time-dependent local diffusivity at the low-temperature regime.** Instantaneous effective diffusivity extracted from the local slope of the MSD traces at 10 K, 30 K, and 60 K. Shaded regions denote the propagated uncertainty from the MSD fits. At 10 K, $D_{eff}$ rises sharply within the first ~50–80 ps, reaching a transient peak of ~25–30 $cm^2$/s before decaying toward near-zero values by ~150 ps. A similar but weaker transient enhancement is observed at 30 K, whereas by 60 K the transport response is largely equilibrated within the experimental window.

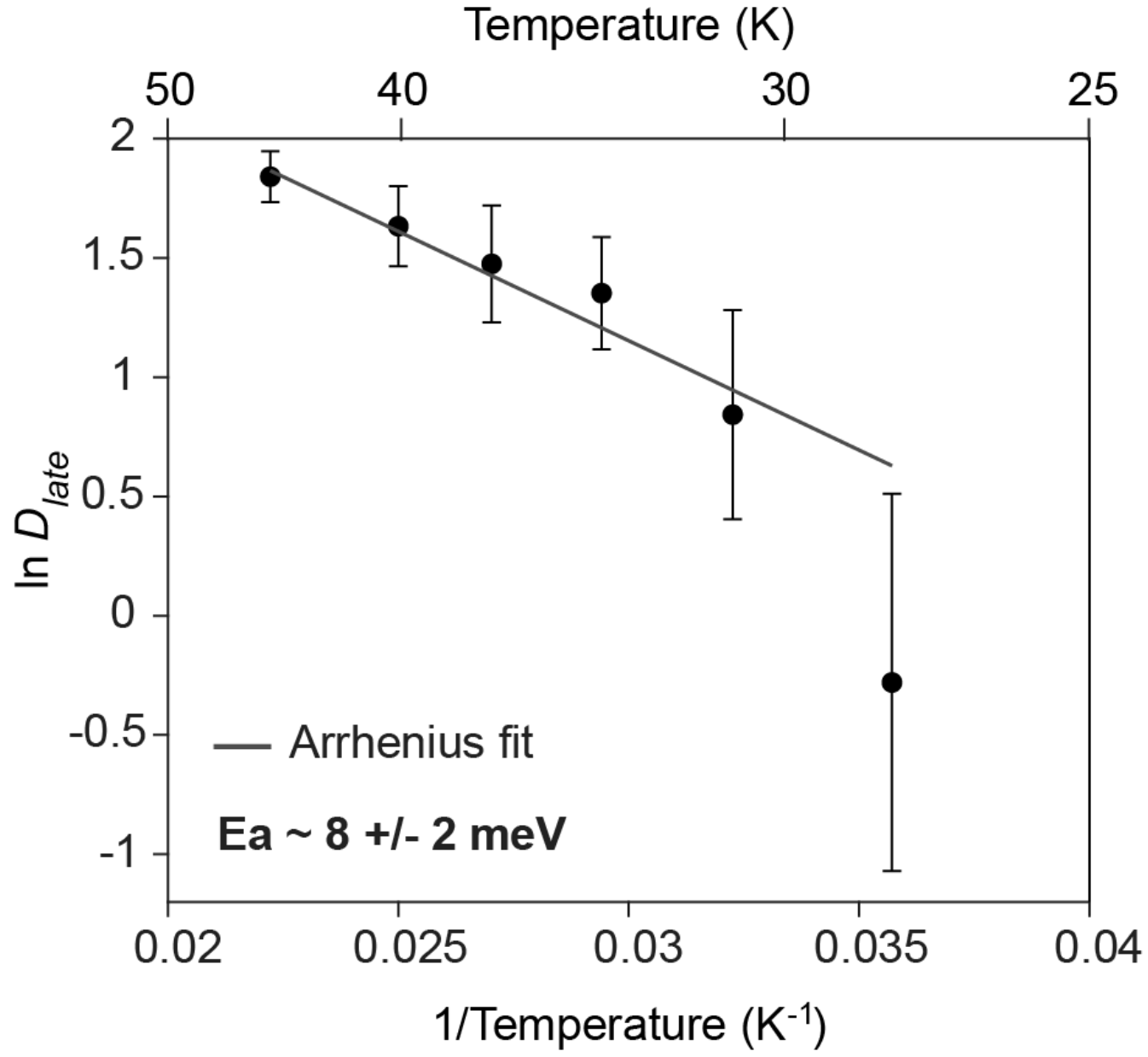

**Figure S12 | Arrhenius plot of $D_{\mathrm{late}}$ for thermally activated transport.** The extracted activation energy $E_a = 8 \pm 2$ meV from the late-time dynamics (>250ps) corresponds to an effective "localization" barrier for thermally activated transport behavior.

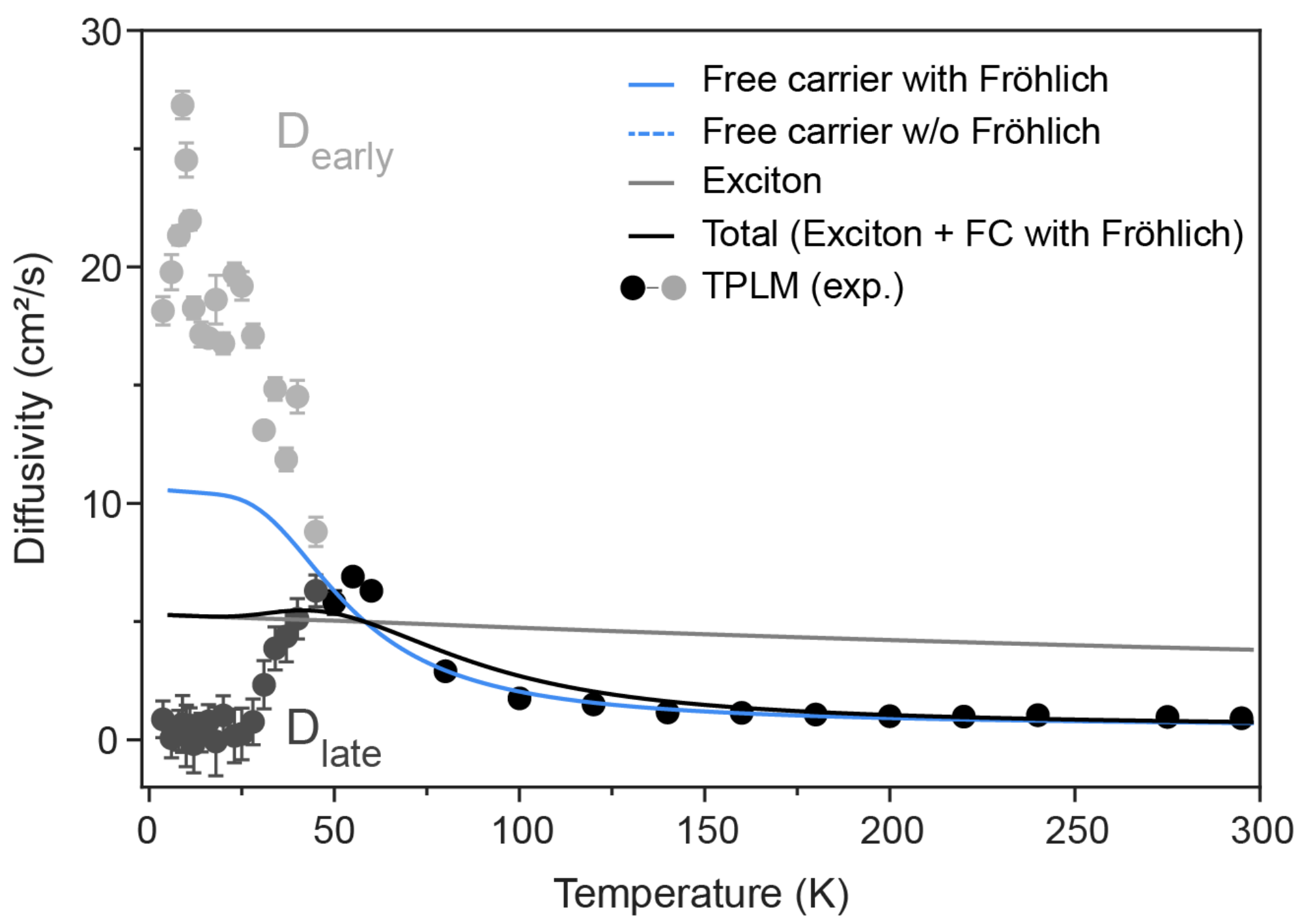

**Figure S13 | Temperature-dependent diffusivity of $CsPbBr_3$, extended comparison between experiment and theory.** We replot the TPLM diffusivities shown in Fig. 3d (main text) together with the full set of theoretical diffusivity curves used to evaluate equilibrium exciton and free-carrier transport. Filled circles show diffusivities extracted from TPLM measurements, while lines show modeled diffusivities for excitons, free carriers with and without Fröhlich coupling, and the Saha-weighted equilibrium mixture. At low temperature, acoustic-phonon scattering dominates, giving nearly temperature-independent diffusivities for the individual exciton and free-carrier branches. In this regime, free carriers are more mobile than excitons because of their smaller effective masses. With increasing temperature, two competing effects shape the total equilibrium response: thermal dissociation increases the free-carrier fraction, while thermally activated optical-phonon scattering suppresses free-carrier mobility. Their competition produces a maximum in the Saha-weighted total diffusivity near 40 K. The experimental fast diffusivities instead increase as temperature decreases, a trend that is not captured by any of the equilibrium model curves. Furthermore, their magnitude remains well above the equilibrium predictions across the cryogenic range, exceeding the Saha-weighted total diffusivity by ~5–6x and the equilibrium free-carrier estimate by ~2–3x. In Section S8 we discuss the origin of this deviation from equilibrium transport models.

## S6. Hot exciton cooling and effective temperature

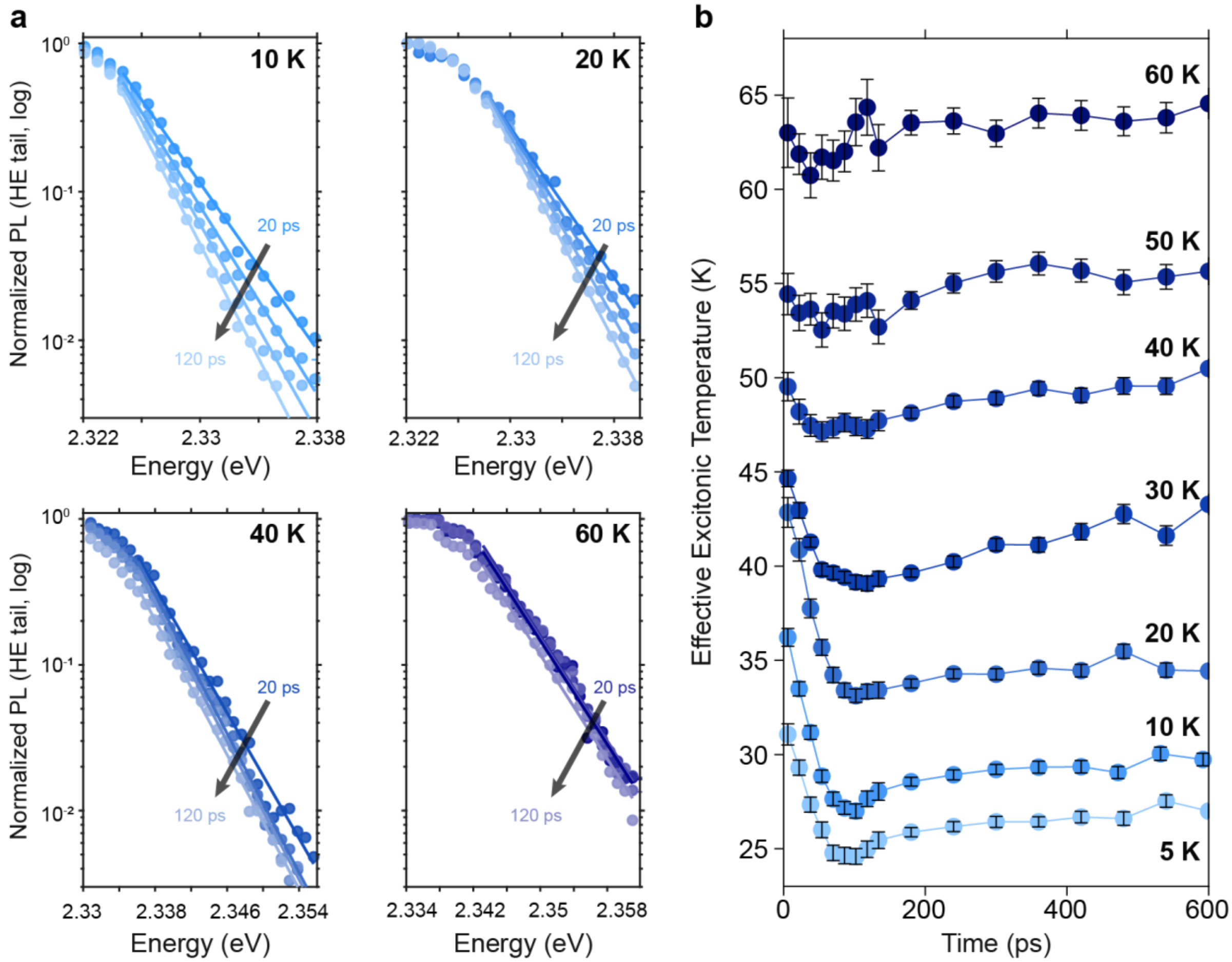

**Figure S14 | Effective excitonic temperature and cooling dynamics (a)** Normalized high-energy (HE) photoluminescence tails plotted on a semilogarithmic scale for selected lattice temperatures at different detection times (20–120 ps). **(b)** Time-dependent effective excitonic temperatures, $T_{eff}$, extracted from the HE tails across 5–60 K. At very low lattice temperatures ($T_{lattice} \approx$ 5-20 K), the transient change in the HE-tail slope shows signatures of cooling dynamics, appearing as a clear decrease in $T_{eff}$ within the first ~100 ps. At later times, however, $T_{eff}$ saturates at an offset value rather than cooling fully to the lattice temperature. The possible origin of this effect is discussed further in Section S6.2. Above 20 K, the cooling signature becomes progressively less pronounced and is nearly absent above ~40 K, where the extracted $T_{eff}$ tracks the lattice temperature from the earliest measured times.

### S6.1 Extraction of effective excitonic temperature

To quantify the cooling of the nonequilibrium exciton population, we extract an effective excitonic temperature, $T_{eff}$, from the high-energy tail of the transient HE photoluminescence at each time delay (Fig. S14). At low lattice temperatures, where the PL linewidth approaches the intrinsic limit of $\Gamma_0$, this spectral region contains contributions from acoustic-phonon-assisted recombination of finite-momentum excitons outside the light cone ($k \neq 0$).[30,31] Its slope therefore provides a measure of the Boltzmann occupation of the hot-exciton distribution. Over the limited high-energy window used here, the tail can be approximated as

$$I(E) \propto \exp\left(-\frac{E - E_0}{k_{\mathrm{B}} T_{\mathrm{eff}}}\right) \tag{6.1}$$

where $E_0$ is a reference energy near the HE peak position. A linear fit of $\ln\ I(E)$ versus $E$ yields a slope $m$, from which an effective temperature can be obtained:

$$T_{\mathrm{eff}} = -\frac{1}{k_B m}. \tag{6.2}$$

We apply this analysis to transient spectra acquired at lattice temperatures from 5 to 60 K (Fig. S14). At each temperature, the fitting window is chosen as the energy range above the HE peak in which $\ln\ I(E)$ remains linear across the analyzed delay times. The energy window is shifted with temperature to follow the HE spectral position while remaining within the exponential tail and is then kept fixed relative to the HE peak for the corresponding time series (Fig. S14a). The robustness of this selection is verified *a posteriori* from the fit diagnostics: across the temperature series, $\ln I(E)$is linear during the cooling period, with representative fits yielding $R^2 > 0.98$ and no systematic residual structure.

### S6.2 Effects of finite emission linewidth

The cooling curves in Fig. S14b reveal a common feature toward the lower lattice temperature limit. $T_{eff}$ saturates at late delays at a value that exceeds $T_{lattice}$, with the offset shrinking as $T_{lattice}$ increases. Before interpreting this offset as evidence of a long-lived hot-exciton population, we consider an alternative origin. The slope method implicitly assumes that the high-energy tail reflects primarily a Boltzmann exponential. In practice, however, the observed PL spectrum is a convolution of the exciton population with the finite homogeneous linewidth. When $k_{\mathrm{B}}T$ becomes comparable to the linewidth, the lineshape itself can dominate the apparent tail, causing the extracted slope to saturate and placing a lower bound on the experimentally resolvable $T_{eff}$.

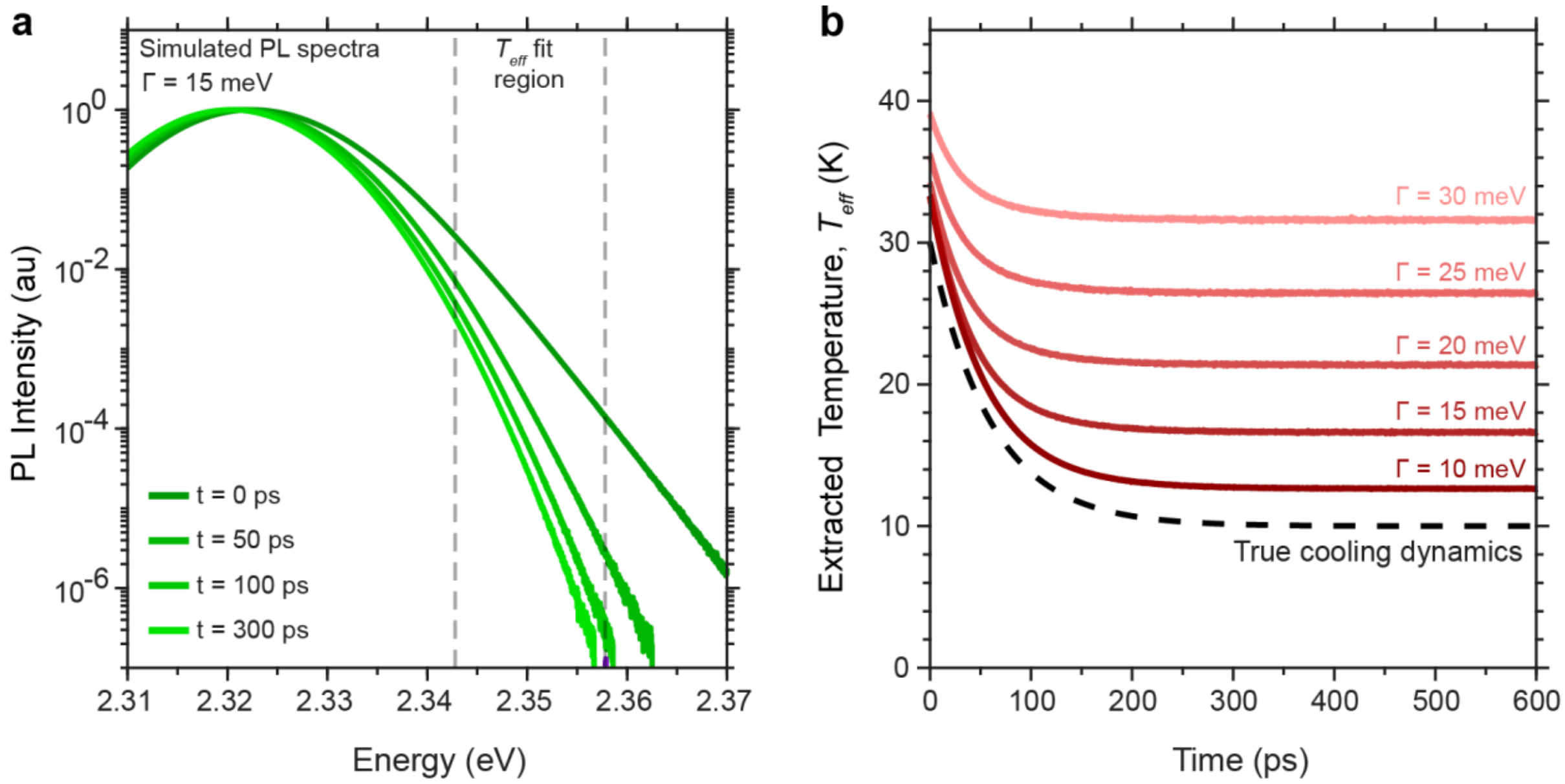

**Figure S15 | Modeling the effect of finite spectral linewidth on apparent cooling dynamics (a)** Simulated emission spectra for a cooling exciton population (cooling parameters of $T: 30 \rightarrow 10$ K, $\tau = 60$ps, chosen for illustration) broadened by a Gaussian of $\mathrm{FWHM} = 15$ meV, comparable to experimentally observed spectral linewidths at low temperatures. Effective temperatures were extracted by fitting the high-energy tail within a window spanning 1.5–2.5 FWHM above the peak position (dashed lines). **(b)** Extracted $T_{\mathrm{eff}}$ obtained from simulated spectra with linewidths spanning $\Gamma = 10–30$ meV. The dashed line indicates the true cooling dynamics from our simulations. Increasing linewidth produces progressively larger deviations between extracted and true lattice temperatures, resulting in an apparent linewidth-dependent temperature floor at long delay times.

---

To assess this effect, we simulated the high-energy emission tail of a cooling exciton population using a minimal model that captures the expected excitonic spectral shape. The intrinsic emission is described by a three-dimensional excitonic density of states weighted by a Boltzmann factor:

$$I(E) \propto \sqrt{E - E_0}\, \exp\left[-\frac{E - E_0}{k_B T}\right], \tag{6.3}$$

This spectrum is then convolved with a Gaussian emission lineshape with FWHM, $\Gamma = 10–30$ meV, spanning the range relevant to the experimentally observed low-temperature linewidths (Fig. S15). Effective temperatures are extracted from the simulated spectra using the same tail-fitting procedure applied to the experimental data, with the fitting window defined relative to the linewidth as 1.5–2.5 FWHM above the late-time peak position.

The simulations show that finite linewidth broadening imposes a lower bound on the experimentally resolvable $T_{eff}$. Even when the underlying exciton population continues to cool, the extracted temperature saturates above the imposed thermal value. Increasing the linewidth produces progressively larger deviations between the extracted and true exciton temperatures, resulting in an apparent linewidth-dependent temperature floor at later detection times. For linewidths comparable to those in our experiment at 10 K (~10-15 meV; Extended Data Fig. 4d), this broadening produces a moderate apparent temperature floor at late times (Fig. S15b). Thus, while the temporal decrease in $T_{eff.}$ provides a robust measure of early hot-exciton relaxation, the absolute late-time value should be interpreted as a linewidth-limited apparent temperature rather than direct evidence for a persistently hot exciton population.

## S7. Two-state kinetic feeding model

One of the key observations of the low-temperature spectral dynamics, beyond the hot exciton effects in the high-energy region, is the delayed rise of the lower-energy (LE) emission relative to the high-energy (HE) feature (Fig. 4c, main text and Extended Data Fig. 6). To capture this behavior, we adopt a minimal two-state kinetic framework comprising a unidirectional feed model. In this picture, photoexcitation initially populates the HE state, while the LE state is unoccupied at $t = 0$. Carriers subsequently relax from HE to LE with a transfer rate constant $k_{\mathrm{tr}}$, and each state undergoes radiative and nonradiative recombination characterized by total decay rates $k_{\mathrm{HE}}$ and $k_{\mathrm{LE}}$. Back-transfer from LE to HE is neglected, consistent with thermal energy <2.6 meV at the low temperatures (<30K) and the energetic separation (~16 meV) between the two emissive states. The resulting population dynamics are therefore described by coupled rate equations for the HE and LE populations, denoted as $N_{\mathrm{HE}}(t)$and $N_{\mathrm{LE}}(t)$, respectively:

$$\frac{dN_{HE}}{dt} = -(k_{HE} + k_{tr})N_{HE}(t) \tag{7.1a}$$

$$\frac{dN_{LE}}{dt} = k_{tr}N_{\mathrm{HE}}(t) - k_{LE}N_{LE}(t) \tag{7.1b}$$

Solving the differential equation (7.1a) yields:

$$N_{HE}(t) = N_0 e^{-(k_{HE}+k_{tr})t} \tag{7.1c}$$

where $N_0$ is the initial HE population. Substituting this expression into Eq. (7.1b) and applying the initial condition $N_{\mathrm{LE}}(0) = 0$ gives:

$$N_{LE}(t) = \frac{k_{tr}N_0}{k_{LE} - (k_{HE} + k_{tr})}\left(e^{-(k_{HE}+k_{tr})t} - e^{-k_{LE}t}\right) \quad (7.1d)$$

This expression shows that the LE population builds up on a timescale governed by the depletion of the HE state ($k_{HE} + k_{tr}$), followed by a decay determined by the intrinsic LE recombination rate ($k_{LE}$). The delayed rise of the LE emission therefore directly reflects the finite transfer time required to populate the LE state.

### S7.1 Global fitting for extracting characteristic timescales

To quantitatively extract kinetic timescales across the full emission spectrum, the transient PL decays were analyzed using a global fitting strategy based on the two-state feed model. The temporal evolution at each emission energy is expressed as a linear combination of shared kinetic basis functions corresponding to the HE population and the LE population, each convolved with the measured instrument response function (IRF). To account for the finite temporal resolution and ensure consistent temporal alignment across all spectral channels, a single global time-zero offset is included as a fit parameter. Only the amplitudes and a constant background are allowed to vary independently with emission energy, while the kinetic parameters are constrained to be global.

The global fit yields two spectrum-independent rate constants,

$$\lambda_1 \equiv k_{HE} + k_{tr}, \lambda_2 \equiv k_{LE},$$

which define the characteristic timescales $\tau_1 = 1/\lambda_1$ and $\tau_2 = 1/\lambda_2$. Here, $\tau_1$ corresponds to the effective depletion timescale of the HE state, incorporating both recombination and transfer, and therefore sets the timescale governing the rise of the LE population. The second timescale, $\tau_2$, describes the intrinsic decay of the LE state once populated.

As a control, we also analyzed the data using a parallel (no-feed) model in which the HE and LE contributions are treated as independent emissive channels with no population transfer between them (see Fig. S16). In this case, the PL dynamics at each energy are described as a linear combination of two IRF-convolved monoexponential decays with independent rates $\lambda_{HE}$ and $\lambda_{LE}$. This model was fit using the same global framework and time-zero alignment as the feed model. In the rest of this section, the resulting kinetic timescales, model comparison metrics, and temperature-dependent fitting behavior are further discussed in Figs. S16–S18 and Tables S1–S2.

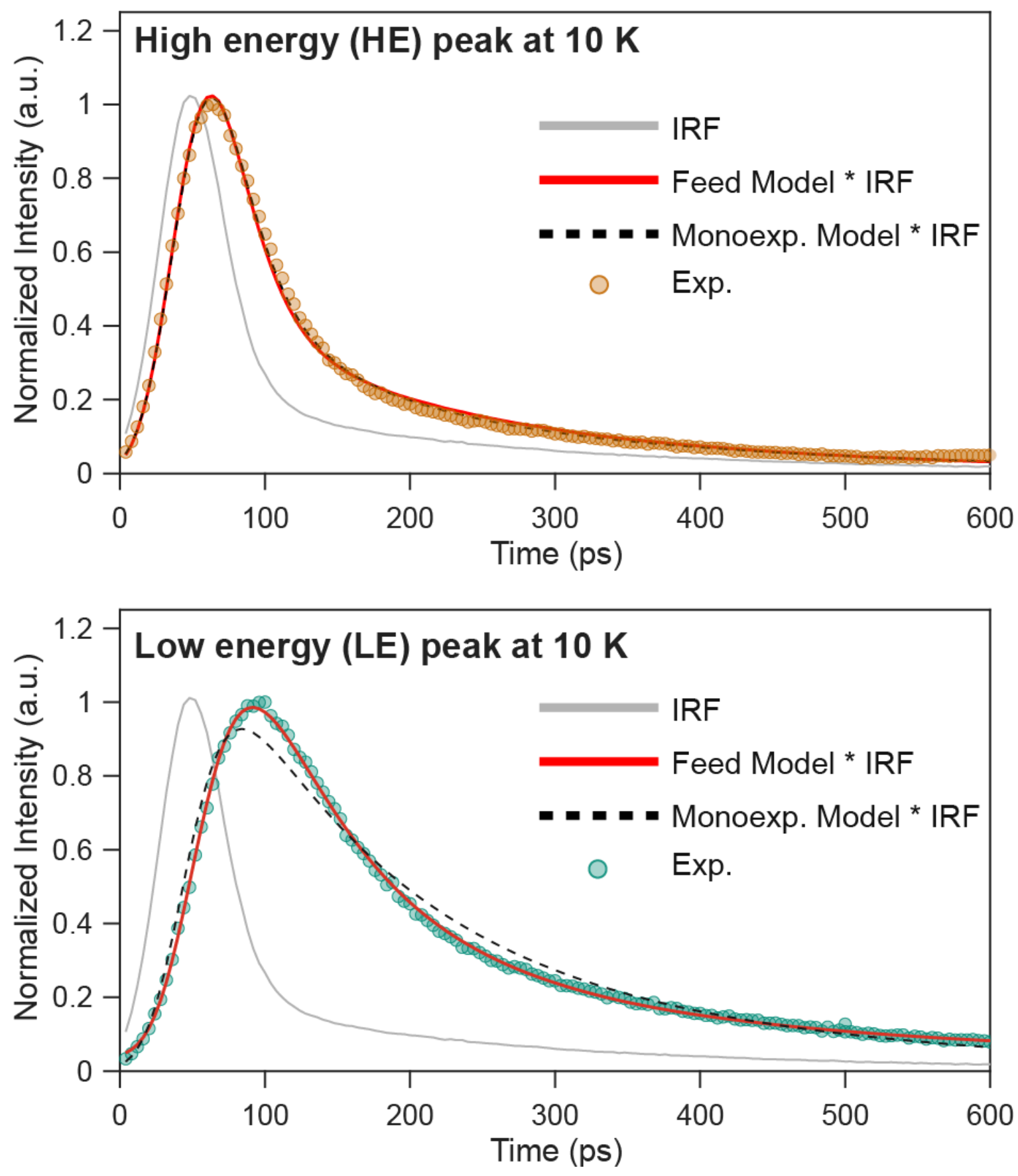

**Figure S16| Model comparison for spectrally-resolved PL dynamics at 10 K.** Normalized PL transients extracted from high-energy (HE, top) and low-energy (LE, bottom) dominated spectral regions at 10 K. Experimental data are compared to the two-state feed model (red solid line) and a parallel no-feed monoexponential decay model (black dashed line), both convolved with the instrument response function (IRF, gray). While both models describe the HE dynamics comparably well as expected, only the feed model captures the delayed rise of the LE emission, which cannot be reproduced by independent decay channels. Model comparison is quantified using the Akaike Information Criterion, where the global difference in AIC, defined as

$$\Delta \mathrm{AIC} = \mathrm{AIC}_{\mathrm{mono}} - \mathrm{AIC}_{\mathrm{feed}} \gg 100$$

shows a decisive statistical preference for the two-state feed model.

---

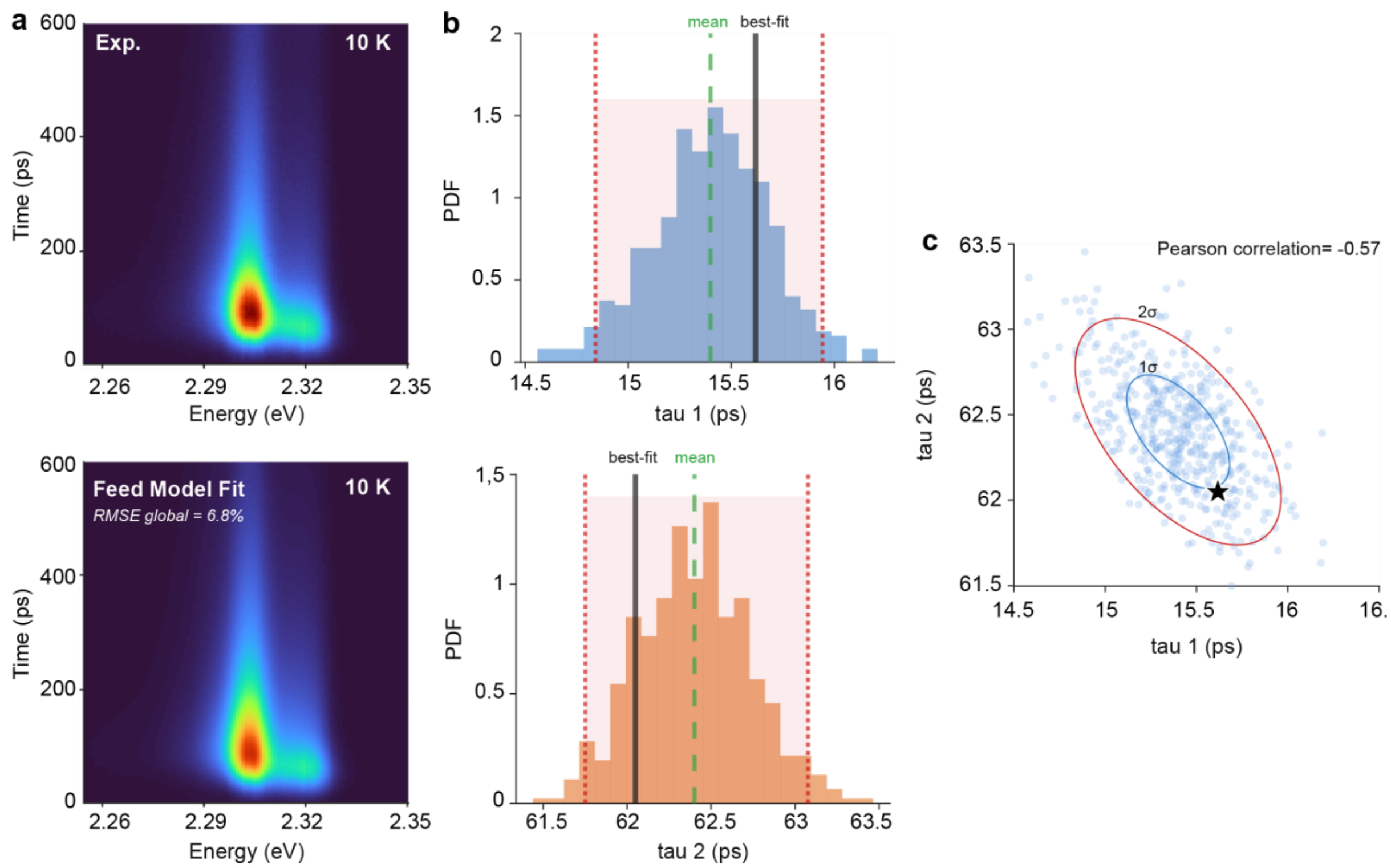

**Figure S17 | Representative feed-model fit at 10 K. (a)** Experimental time-resolved PL intensity energy-map at 10 K (top) and corresponding global two-state feed model fit**.** The model captures the early-time high-energy emission and the subsequent population transfer into the lower-energy state, with a global RMSE 6.8% (**b)** Bootstrap distributions of the extracted lifetimes $\tau_1$(top) and $\tau_2$(bottom). The black solid line indicates the best-fit value, the green dashed line the bootstrap mean, and red dotted lines the 95% confidence interval. Both distributions are narrow, indicating strong overall constraint of the kinetic parameters, with a slight asymmetry leading to an offset between the best-fit and mean values. The corresponding timescales are $\tau_1 = 15.7^{+0.3}_{-0.8}$ ps and $\tau_2 = 62.0^{+1.1}_{-0.2}$ ps. **(c)** Joint bootstrap distribution of $\tau_1$and $\tau_2$, showing anticorrelation (Pearson $r = -0.57$), indicating that while the parameter pair is tightly localized, the individual lifetimes are not fully independent. Fit results for additional temperatures are summarized in Table S1.

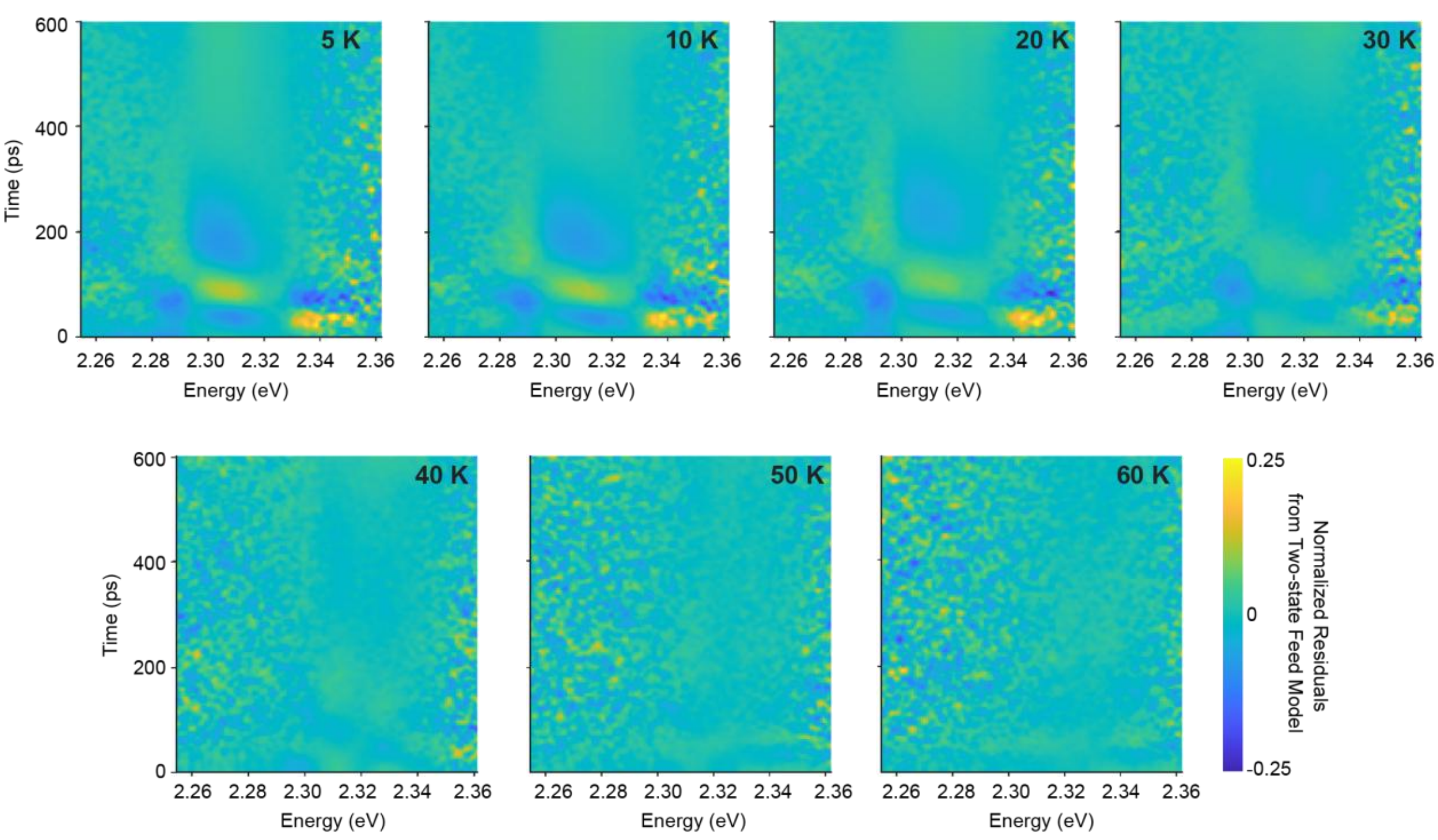

**Figure S18 | Residual maps from the two-state global feed model.** Residuals are defined as the difference between the normalized experimental time-resolved PL and the corresponding global fit, evaluated over the same energy-time window and displayed on a common color scale. Overall the two-state feed model captures the main spectral evolution across the temperature series. At higher temperatures (≥40 K), residuals are weak and largely featureless, and the model provides a good description of the spectral dynamics in this regime. At low temperatures (5–20 K), the model still captures the overall spectral evolution, but pronounced structured residuals emerge. Strong positive residuals appear at early times (<150 ps) in the high-energy region (2.33–2.35 eV), where the model underestimates the initial high-energy population. Weaker, subsequent deviations near the main emission band (~2.31 eV) likely reflect downstream consequences of this nonequilibrium initial distribution. Together, these features are consistent with emission from a transient population of excitons carrying excess kinetic energy, whose dynamics the static two-state framework — with its fixed high-energy amplitude profile and single kinetic feed — cannot fully accommodate. Importantly, the temperature range over which this residual structure emerges (<40 K) coincides with the regime in which transport deviates from diffusive behavior (Fig.3c, main text) and cooling dynamics emerge (Fig.4b, main text), linking the deviations from the feed model to the broader emergence of nonequilibrium exciton dynamics.

**Table S1 | Two-state feed-model fitting parameters and statistics as a function of temperature.** Extracted timescales $\tau_1$ and $\tau_2$, global RMSE and Pearson correlation coefficient between the two parameters. Across all temperatures, ΔAIC$\gg$ $100$ values show decisive statistical preference for the feed model over a parallel no-feed model.

| Temp. (K) | $\tau_1$ **(ps)** | $\tau_2$ **(ps)** | global RMSE (%) | Pearson Correlation |
|---|---|---|---|---|
| 5 | $13.2\,^{+0.4}_{-0.8}$ | $70.5\,^{+1.2}_{-0.3}$ | 6.1 | -0.33 |
| 10 | $15.7\,^{+0.3}_{-0.8}$ | $62.0\,^{+1.1}_{-0.2}$ | 6.8 | -0.57 |
| 20 | $27.4\,^{+0.4}_{-0.7}$ | $86.2\,^{+1.1}_{-0.3}$ | 7.2 | -0.52 |
| 30 | $37.7\,^{+0.9}_{-0.6}$ | $127\,^{+2.0}_{-1.0}$ | 6.7 | -0.59 |
| 40 | $45.9\,^{+1.0}_{-0.4}$ | $252\,^{+3.0}_{-1.0}$ | 5.0 | -0.49 |
| 50 | $68.8\,^{+2.3}_{-0.4}$ | $307\,^{+3.0}_{-2.0}$ | 4.2 | -0.22 |
| 60 | $90.5\,^{+2.1}_{-1.0}$ | $552\,^{+6.0}_{-4.0}$ | 5.2 | -0.30 |

The extracted kinetic parameters from the two-state feed model are summarized in Table S1. The temperature dependence of the resulting timescales provides direct insight into the nature of the LE emissive state. Both $\tau_1$ and $\tau_2$ increase strongly with temperature, from 13 to 90 ps and from 70.5 to 552 ps, respectively. Since transfer dominates over direct HE recombination at low temperatures (see Section S7.2), the increase in $\tau_1$ primarily reflects a progressive reduction in transfer efficiency into the LE state. Simultaneously, the pronounced increase in $\tau_2$ implies that the LE population becomes increasingly long-lived with temperature. Collectively, these trends support a picture in which the LE state behaves as a shallow trap, such that as $k_B T$ approaches the trapping depth, the HE and LE populations move toward thermal equilibrium. This may reduce the net capture rate while allowing thermal detrapping and recapture processes to progressively contribute to the observed increase in both timescales. Consistent with this interpretation, a phenomenological Arrhenius analysis of $\tau_2$ yields an activation energy of $E_a = 7.0 \pm 1.1$ meV, comparable to the activation energy extracted from the late-time TPLM transport dynamics (Fig. S12).

This picture is supported by the time-averaged PL spectra (Extended Data Fig. 6), where the LE feature is well resolved at 5–20 K but progressively merges with the main emission by ~40 K, consistent with thermal depopulation of a shallow trapped state. At the lowest temperatures, capture is efficient and effectively irreversible, yielding a short $\tau_1$ and a $\tau_2$ that more closely reflects the intrinsic LE lifetime. At elevated temperatures, capture becomes less efficient and partially reversible, producing a longer $\tau_1$ and an apparent increase in $\tau_2$ through carrier recycling between the two states.

### S7.2 High-energy state recombination and transfer contributions

As discussed above, the global analysis yields two characteristic timescales, $\tau_1^{-1} = k_{\mathrm{HE}} + k_{\mathrm{tr}}$ and $\tau_2^{-1} = k_{\mathrm{LE}}$, such that the HE dynamics reflect a combined depletion rate and therefore do not independently determine $k_{\mathrm{HE}}$ and $k_{\mathrm{tr}}$. However, at low temperatures (5–20 K), where $k_B T \lesssim$ 1.7 meV remains much smaller than the HE–LE energetic separation (~ 14–16 meV), back-transfer is likely suppressed. In this limit, and assuming comparable radiative yields with nonradiative losses small compared with the radiative and transfer channels, the integrated intensity ratio between the two emissive states provides an estimate of the branching between direct HE recombination and transfer into the LE state:

$$\frac{I_{LE}}{I_{HE}} \approx \frac{k_{tr}}{k_{HE}} \tag{7.2}$$

Using $I_{\mathrm{LE}}: I_{\mathrm{HE}} = 1: 0.3$ at 10 K results in $\tau_{\mathrm{HE}} \approx \mathbf{68}$ **ps** and $\tau_{\mathrm{tr}} \approx \mathbf{20}$ **ps**. Applying the same estimation across 5–20 K (Table S2 below) yields consistently $k_{\mathrm{tr}}/k_{\mathrm{HE}}$ ~3-3.5, indicating that inter-state transfer dominates over direct HE recombination throughout this temperature range ($k_{\mathrm{tr}} > k_{\mathrm{HE}}$). Thus, the apparent fast HE lifetime in our spectral dynamics (Extended Data Fig.6) primarily reflects inter-state transfer rather than intrinsic HE recombination.

At higher temperatures (e.g., 30 K), the HE feature is no longer clearly resolved in the time-averaged spectra, the estimation is therefore restricted to the 5–20 K regime.

**Table S2 | Estimated HE recombination ($\tau_{\mathrm{HE}}$) and HE→ LE ($\tau_{\mathrm{tr}}$ ) transfer contributions.**

| *Temp. (K)* | $\tau_{HE}$ *(ps)* | $\tau_{tr}$ *(ps)* |
|---|---|---|
| *5* | 57 | 17 |
| *10* | 68 | 20 |
| *20* | 120 | *36* |

## S8. Why do hot excitons diffuse fast?

Diffusion is determined by the interplay between quasiparticle group velocity and scattering, as shown by Eqs. (1.5) and (1.6) for carriers and excitons (see section S1). At cryogenic temperatures, the scattering time depends weakly on momentum, $\tau_{\mathbf{q}}^{a} \approx \tau^{a}$, because thermal occupation of states energetic enough to emit optical phonons is strongly suppressed. As a consequence, within this approximation the diffusion coefficient can be written as:

$$D_a \approx \frac{2}{3M_a} \tau^a \langle E \rangle_a , \tag{8.1}$$

Here, $M_a$ is the quasiparticle effective mass, $\tau^a$is the momentum-scattering time, and $\langle E \rangle_a = \sum_{\mathbf{q}} E_{a;\mathbf{q}} f_{\mathbf{q}}^{e/h,\circ}$ is the average kinetic energy of the distribution, with $E_{a;q} = \hbar^2 q^2 / 2M_a$. Equation (8.1) therefore shows that diffusivity increases when quasiparticles carry more kinetic energy, with similar arguments applying for excitons and carriers ($a$ =X,e,h).

At equilibrium, one has the average kinetic energy as $\langle E \rangle_a = \frac{3}{2} k_B T$ as per equipartition theorem, which increases linearly with temperature. In the low-temperature limit considered here, scattering is dominated by acoustic phonons, for which $\tau^a \approx \hbar / (c_{\mathrm{ac}} T)$. The expected equilibrium diffusion can then be written as:

$$D_a = \frac{\hbar k_B}{c_{ac} M_a} \tag{8.2}$$

The consequence of Eq. 8.2 is that, within the acoustic-phonon-limited equilibrium picture, the diffusion coefficient of a given species is expected to become nearly temperature independent at low temperatures, as shown by the theoretical diffusion curves in Fig. S13.

This equilibrium picture changes when the exciton population is formed out of equilibrium. Here the average kinetic energy entering the diffusivity relation is set by the excess energy $E_{\mathrm{ex}}$ of the hot exciton population. For thermal energies $k_B T \ll E_{ex}$, the resulting nonequilibrium diffusion coefficient reads: [32]

$$D_{a;neq} = \frac{\hbar E_{ex}}{c_{ac} T M_a} \tag{8.3}$$

This implies that the transient diffusivity increases with decreasing temperature, in contrast to what is expected at equilibrium and in good agreement with our experiments. In Extended Data Fig. 7a, we calculate this transient diffusivity for several excess energy values. Within the kinetic-bottleneck picture, one might expect hot excitons to be formed with excess energies on the order of the exciton binding energy, $E_b \approx 19$ meV, if the continuum-to-exciton relaxation releases a substantial fraction of this energy into exciton kinetic motion. However, calculations using such large excess energies overestimate the experimentally observed transient diffusivities.

This comparison suggests that the emissive hot-exciton population probed experimentally carries a much smaller effective excess energy than the full binding energy. The measured diffusivities are instead more consistent with $E_{\mathrm{ex}} \sim 2\text{-}3$ meV (Extended Data Fig. 7a). This reduction is physically reasonable because the analytical estimate assumes that excitons are formed with a prescribed excess energy and that their motion is directly resolved. In the experiment, hot-exciton formation occurs over a finite timescale, so carriers may already partially cool during continuum-to-exciton transfer— and a fraction may already relax toward the lower-energy LE state— resulting in reduced early-time diffusivities. Additional relaxation through low-energy optical phonons can further reduce the kinetic energy retained by the emissive exciton population. Moreover, the finite instrument response function temporally averages the earliest and fastest part of the spatial expansion, lowering the apparent diffusivity extracted from the measured PL dynamics. Thus, while excess energies comparable to $E_b$ provide an upper-bound scenario, the experimentally observed transport points to a cooler nonequilibrium exciton population within our experimental window. The simplified treatment nevertheless captures the central qualitative feature of the low-temperature data, that finite excess kinetic energy can produce a transient diffusivity that decreases with increasing temperature.

A second consequence of finite excess energy is that it favors band-like transport by moving the exciton population farther from the localization threshold associated with quantum-interference effects.[12,13] This localization can take place when the mean-free-path $\lambda_{\mathrm{mfp}} = \sqrt{2\mathrm{D}_{\mathrm{X;neq}}\tau_{\mathrm{X}}}$ of the exciton becomes similar to or smaller than the de Broglie wavelength $\lambda_{dB} = \frac{\hbar 2\pi}{\sqrt{2M_X(\frac{3}{2}k_B T + E_{ex}}}$, as described by the Mott-Ioffe-Regel (MIR) limit. In Fig. 4d & Extended Data Fig.7b in the main text, we have looked at the ratio between those two quantities as a function of temperature and excess energy (taking a temperature-independent exciton mass $M_X = 0.4m_0$). In general, the presence of an excess energy results in a larger ratio, thus indicating band-like transport. Only in the presence of negligible excess energies does such a ratio go below one, indicating the possibility of exciton localization. [12,13]

## References for SI